\documentclass[aps,prab,reprint,preprintnumbers]{revtex4-2}

\RequirePackage{graphicx}
\usepackage{GRAF}
\newcommand{\tten}[1]{\!\cdot\! 10^{#1}}
\newcommand{\Ifpc}{I_{\mathrm{FPC}}}
\newcommand{\Ncit}{N_{\mathrm{CIT}}}
\newcommand{\Qext}[1]{Q_{\mathrm{ext,#1}}}

\RequirePackage[colorlinks,linkcolor=blue,urlcolor=blue,bookmarksnumbered=true]{hyperref}
\RequirePackage[figure]{hypcap}
\RequirePackage[capitalise,nameinlink]{cleveref}
\RequirePackage{ifthen}

\begin{document}

\title{Report on first plasma processing trial for a FRIB quarter-wave resonator
cryomodule}

\author{Walter~Hartung}
\author{Wei~Chang}
\author{Yoo-Lim~Cheon}
\author{Kyle~Elliott}
\author{Sang-Hoon~Kim}
\author{Taro~Konomi}
\author{Patrick~Tutt}
\author{Yuting~Wu}
\author{Ting~Xu}
\affiliation{Facility for Rare Isotope Beams, Michigan State University, East Lansing, MI, USA}
\date{10 November 2025}

\begin{abstract}
  Plasma processing has been shown to help mitigate degradation of the
  performance of superconducting radio-frequency cavities, providing
  an alternative to removal of cryomodules from the accelerator for
  refurbishment.  Studies of plasma processing for quarter-wave
  resonators (QWRs) and half-wave resonators (HWRs) are underway at
  the Facility for Rare Isotope Beams (FRIB), where a total of 324
  such resonators are presently in operation.  Plasma processing tests
  were done on several QWRs using the fundamental power coupler (FPC)
  to drive the plasma, with promising results.  Driving the plasma
  with a higher-order mode allows for less mismatch at the FPC and
  higher plasma density.  The first plasma processing trial for 
  FRIB QWRs in a cryomodule was conducted in January 2024.  Cold
  tests of the cryomodule showed a significant reduction in field
  emission X-rays after plasma processing.
\end{abstract}

\maketitle

\tableofcontents

\section{Introduction}

\newlength{\landwidth}%
\newlength{\sqwidth}%
\newlength{\triwidth}%
\newlength{\portwidth}%
\ifthenelse{\lengthtest{\columnwidth=\textwidth}}
{
\setlength{\landwidth}{0.75\columnwidth}%
\setlength{\sqwidth}{0.6\columnwidth}%
\setlength{\triwidth}{0.5\columnwidth}%
\setlength{\portwidth}{0.45\columnwidth}%
}%
{
\setlength{\landwidth}{\columnwidth}%
\setlength{\sqwidth}{\columnwidth}%
\setlength{\triwidth}{\columnwidth}%
\setlength{\portwidth}{\columnwidth}%
}%

Superconducting radio-frequency (SRF) cavities provide high
accelerating gradients with low power dissipation for modern particle
accelerators, but they cannot always operate in the best possible
environment and do not always reach the best possible performance.
Plasma processing has been developed and widely adopted by the
electronics industry for modification of surfaces
\cite{LIEBLICHT2005:PrincPlasDisMatProc}.  Plasma processing has been
applied to SRF cavities in recent years, with the goal of improving SRF
cavity performance or reversing performance degradation.  In-situ
plasma processing provides an alternative to time-intensive and
labor-intensive removal and disassembly of cryomodules for
refurbishment of each cavity via repeat chemical etching and rinsing.

Pioneering work at the Spallation Neutron Source (SNS) showed that
plasma processing can be beneficial for an SRF accelerator
\cite{LINAC2016:WE2A03, NIMA852:20to32}.  SRF-cavity plasma processing studies are now
underway at several other facilities
\cite{TTC:JUN2018:BERRUTTI:PLASMA, NIMA893:95to98, LINAC2016:MOPRC029, NIMA905:61to70,
  SRF2019:FRCAB6, SRF2019:FRCAB7, PRAB24:022002, PRAB25:102001,
  SRF2023:MOPMB049, SRF2023:THIXA02,
  TTC:DEC2023:CHENEY:PLASMA, IPAC2024:WEPS52, LINAC2024:THXA005,
  LINAC2024:THPB089, PRAB27:123101}.  Notable recent results include
in-cryomodule plasma processing for LCLS-II \cite{PRAB25:102001} and
in-tunnel plasma processing work at CEBAF \cite{IPAC2024:WEPS52, LINAC2024:THXA005},
both further demonstrating performance improvement with plasma
processing.  In addition to the work on multi-cell elliptical SRF
cavities for SNS, CEBAF, LCLS-II \cite{SRF2019:FRCAB7, PRAB24:022002,
  PRAB25:102001}, and on ILC-type cavities
\cite{NIMA893:95to98, LINAC2024:THPB089, PRAB27:123101}, plasma processing is being
studied for quarter-wave resonators (QWRs) at IJCLab
\cite{TTC:DEC2023:CHENEY:PLASMA}, half-wave resonators (HWRs) at IMP
\cite{LINAC2016:MOPRC029, NIMA905:61to70, SRF2019:FRCAB6} and Argonne
\cite{SRF2023:MOPMB049}, and spoke resonators at Fermilab
\cite{TTC:JUN2018:BERRUTTI:PLASMA, SRF2023:THIXA02}.

The Facility for Rare Isotope Beams (FRIB) is an SRF linac which
accelerates ions to $\geq 200$~MeV per nucleon; user operations began
in May 2022 \cite{MPLA37:2322300063}.  The beam power ramp-up from
1~kW for initial user experiments to the ultimate performance goal of
400~kW is ongoing~\cite{JINST19:T05011}.  As the linac contains 104
QWRs (80.5 MHz) and 220 HWRs (322 MHz) \cite{NIMA1014:165675}, the
risk of cavity performance degradation due to contamination is a
concern for long-term operation.  This concern let us to initiate a
plasma processing development effort for FRIB cavities in 2020.  This
paper summarizes the QWR development efforts and presents the first
results of off-line plasma processing of a spare FRIB QWR cryomodule,
with before-and-after cold tests of the cryomodule to assess the
impact of plasma processing on cryomodule performance.

\section{FRIB cavities and cryomodules}

FRIB production cavities were fabricated from sheet niobium; jacketed
cavities were delivered by industrial suppliers \cite{SRF2015:WEBA03,
  SRF2017:TUXAA03}, and then cold tested at FRIB
\cite{NIMA1014:165675} after buffered chemical polishing (BCP), and
high-pressure water rinsing (HPWR\@).  Certified cavities were
assembled into cryomodules \cite{SRF2017:FRXAA01} and cold tested at
FRIB \cite{NAPAC2019:WEPLM73} before installation into the linac
tunnel.  RF pulse conditioning has been used to mitigate field
emission during cryomodule bunker testing and in the linac
tunnel~\cite{LINAC2024:TUPB013, HPFOUR:PAPER:CHEON}.

Additional spare cryomodules are presently being fabricated.  The
first spare $\beta_m = 0.086$ QWR cryomodule was used for the first
FRIB cryomodule plasma processing trial ($\beta_m$ = optimum
normalized beam speed $v/c$).  The cryomodule contains 8 cavities
interleaved with 3 superconducting solenoids, all operating at
4.5~K~\cite{SRF2019:WETEA5}.  Several of the cavities for the spare
cryomodule were tested more than once, with iteration on BCP and HPWR
in an effort to reduce field emission below the onset levels seen
during FRIB cryomodule production.  All of the cavities had a
low-temperature bake-out prior to their final Dewar test, a step which
was not routinely included during FRIB cryomodule production.

\section{Plasma development stages}

Plasma processing is done with the cavities at room temperature.  Our
end goal is to develop the capability for plasma processing of FRIB
cryomodules in the tunnel.  We used a step-wise approach
\cite{SRF2021:WEPTEV011, NAPAC2022:MOPA91, SRF2023:THIXA01}.  Major
challenges for FRIB cavities are (i) the fundamental power coupler
(FPC) mismatch at room temperature, (ii) the absence of higher-order
mode (HOM) couplers, and (iii) the limited view of the cavity interior
through access ports.  These raise concerns about ignition of plasma
in the FPC rather than in the cavity, which would risk damage to the
FPC or cavity.

The use of a higher-order mode is helpful in reducing mismatch and
mitigating the risk of coupler ignition.  HOMs have been found useful
for the LCLS-II \cite{PRAB25:102001} and CEBAF \cite{IPAC2024:WEPS52, LINAC2024:THXA005}
cavities, both of which have more FPC mismatch at room temperature
than the SNS cavities.  The plasma ignition threshold RF field
increases in an approximately linear manner with RF frequency
\cite{SRF2023:WEPWB127}, so the RF power required for plasma ignition
using HOMs may be larger than that needed for the fundamental mode,
even though the matching is improved.

\begin{figure}[hbp]
\includegraphics[width=\columnwidth]{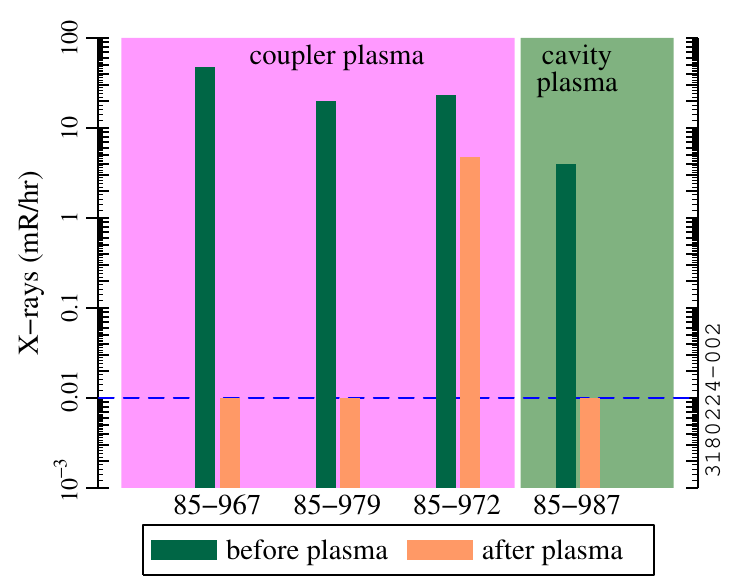}
  
\caption{Measured field emission X-rays at $E_a = 10$~MV/m in cold
  tests of FRIB $\beta_m = 0.086$ QWRs before and after ``on-the-bench''
  plasma processing with an FPC set for maximum
  coupling.  The dashed line indicates the X-ray sensor
  background level.\label{F:xbench}}
\end{figure}

Most of the development work was done using a neon-oxygen plasma at
$\sim 100$~mTorr.  First plasma trials were done with the fundamental
mode, using custom-length input couplers to provide near-unity coupling
at room temperature, \cite{SRF2021:WEPTEV011}.  Subsequently, three
FRIB QWRs were plasma-processed using the FPC and the TEM $5\lambda/4$
mode \cite{NAPAC2022:MOPA91}.  However, in follow-up measurements, we
determined that the plasma was in the coupler region rather than in
the cavity, even though a reduction in cavity field emission (FE) was
seen after coupler plasma processing \cite{SRF2023:THIXA01}; X-ray
measurements at an accelerating gradient ($E_a$) of 10 MV/m in
before-and-after cavity tests are shown in \cref{F:xbench} (pink
background).  In two early trials with FRIB HWRs, sputtering of copper
from the custom-length input antenna onto the niobium RF port was
observed after coupler plasma ignition \cite{SRF2021:WEPTEV011,
  SRF2023:THIXA01}.

After better differentiation between cavity and coupler plasma, plasma
processing on a QWR and an HWR was done with cavity plasma, shifting
the drive frequency up in order to produce a more dense plasma; an
improvement in FE X-rays was observed for both
\cite{SRF2023:THIXA01}; the measured X-rays for the QWR are included
in \cref{F:xbench} (green background).  Note that the gradient of
$10$~MV/m used as a reference field in \cref{F:xbench} is well
above the gradient goal for FRIB Linac operations ($E_a = 5.6$~MV/m).

One of the plasma-processed cavities (S85-967) was installed into the
spare QWR cryomodule.  We plan to use the remaining plasma-processed
cavities in future cryomodules.

\section{Plasma processing system and methods}

\subsection{Coupler position}

\begin{figure}[bp]
\includegraphics[width=\columnwidth]{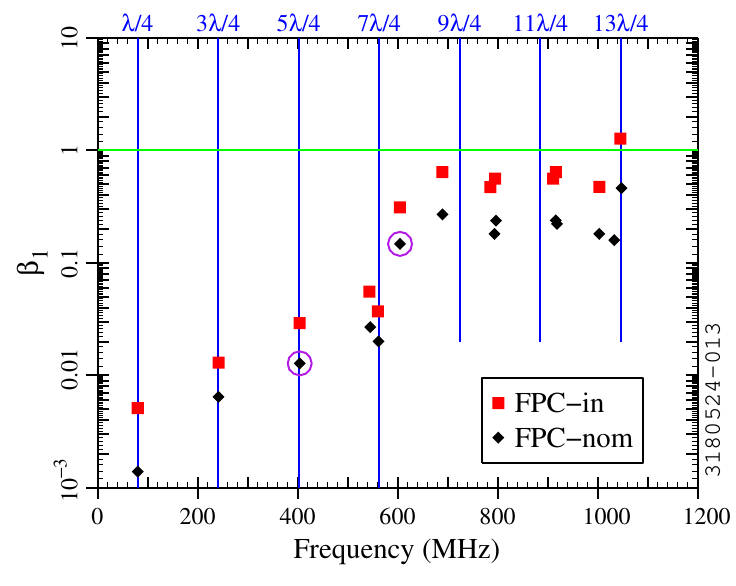}
  
\caption{Measured FPC coupling factor ($\beta_1 = Q_0/\Qext{1}$) as a
function of frequency for some of the modes in a FRIB
$\beta_m = 0.086$ QWR, with the FPC set for maximum coupling (red
squares) or nominal coupling (black diamonds).\label{F:vna}}
\end{figure}

The FPCs for the FRIB QWRs allow for manual coupling adjustment
\cite{NAPAC2016:WEB3IO01}, which is straightforward for
``on-the-bench'' plasma processing.  Hence, the FPC was set for 
stronger coupling in the plasma development stage.  However,
adjustment of the FPCs in a QWR cryomodule requires venting of the
insulating space, removal of access ports, and partial disassembly of
the thermal shield and multi-layer insulation.  Consequently, we
transitioned to using the nominal FPC position before starting plasma
processing on a QWR cryomodule.  As seen in \cref{F:vna}, this
produces more mismatch, making plasma processing more challenging.
With the FPC in nominal position, we estimate that the maximum plasma
density is about 1/3 of the density that can be reached with the FPC
adjusted for stronger coupling (based on bench studies with a Ne/O$_2$ plasma).

\subsection{Drive modes}

\begin{figure}
\newlength{\imgheight}%
\setlength{\imgheight}{0.435\textheight}%
\raisebox{0.9\imgheight}{\Large (a)} %
\includegraphics[height=\imgheight]{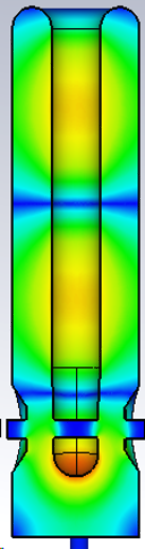} %
\includegraphics[height=\imgheight]{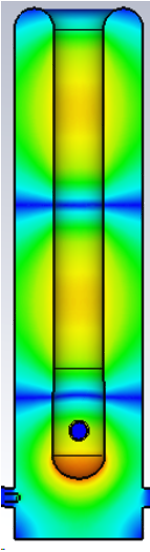}\\
\raisebox{0.9\imgheight}{\Large (b)} %
\includegraphics[height=\imgheight]{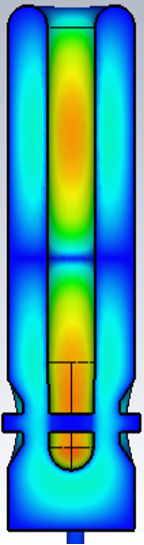} %
\includegraphics[height=\imgheight]{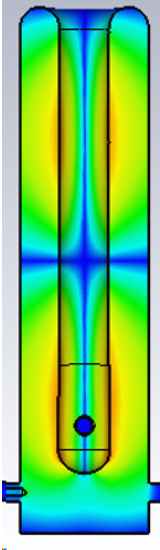} %

\caption{Sectional views of the FRIB $\beta_m = 0.086$ QWR with
  intensity maps (blue = low, red = high) of the electric field
  magnitude for (a) the 404~MHz TEM mode and (b) the 605~MHz dipole
  mode.  The side views (left) show the beam ports and drift tube.
  The front (``beam's eye'') views (right) show the RF ports for the
  input/FPC and pickup probe couplers below the beam
  line.\label{F:maps}}

\end{figure}

As seen in \cref{F:vna}, there is less mismatch for higher-frequency
modes.  Two HOMs (circled in purple in \cref{F:vna}) were selected to
drive the plasma for the cryomodule trial, the TEM $5\lambda/4$ mode
($\sim 404$~MHz) used in previous bench trials, and the second dipole
mode ($\sim 605$~MHz).  Fields from CST Microwave Studio
\cite{ICAP2006:THM2IS03} are shown in \cref{F:maps}.  One would expect
high plasma density where the electric field is high, though we
observe that the plasma tends to ignite in only one high-field region,
which is a disadvantage for HOMs.  Previous studies showed analogous
behavior in multi-cell elliptical cavities: if the TM$_{010}$~$\pi$
mode is used with the hope of producing a uniform plasma throughout
the cavity, the plasma tends to ignite in only one cell
\cite{JAP120:243301}.

The 404~MHz TEM mode has a favorable plasma distribution, as the
plasma consistently ignites in the bottom of the three high-electric
field lobes, which provides good coverage of the surfaces likely to
produce field emission in operation; however, the coupler mismatch
does not allow for very high plasma density (limited by the onset of
coupler ignition).  This mode still allows for a plasma density more
than 300 times higher than that of the fundamental mode (in bench
studies with a Ne/O$_2$ plasma).  The preferential formation of plasma
in the bottom lobe is likely due to the higher electric field at the
lower tip of the inner conductor for the 404~MHz mode
(\cref{F:maps}a).

The 605~MHz dipole mode has a less favorable plasma distribution, with
plasma likely present in only 1/4 of the cavity, based on bench
studies.  On the other hand, higher plasma density can be reached
(limited by the available RF power).  We chose this mode because it
provided the highest plasma density we could achieve using the same RF
components as for the 404~MHz mode.

\subsection{System and interlocks}

\begin{figure}
\includegraphics[width=\columnwidth]{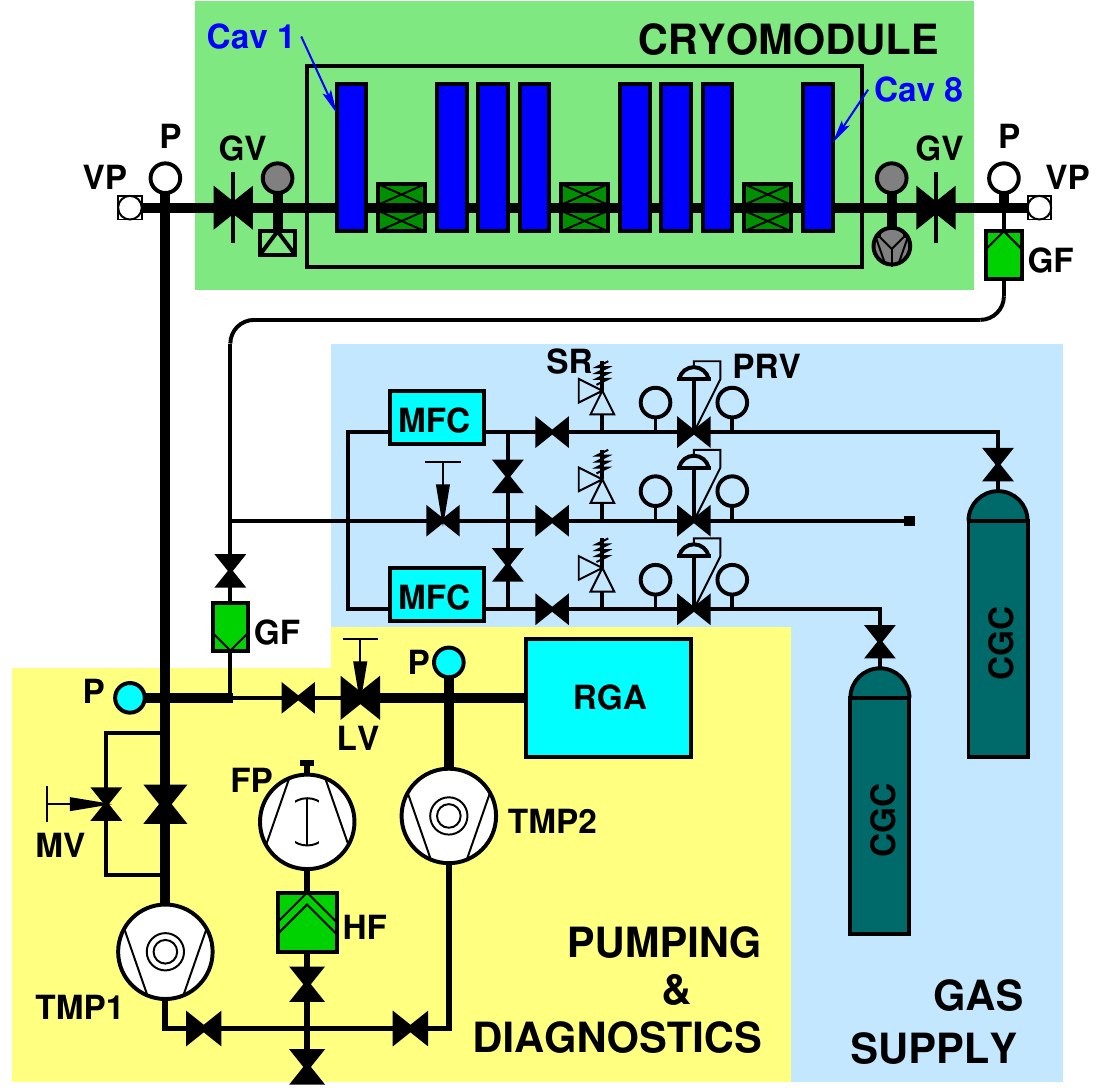}
  
\caption{Schematic of the gas supply and pumping system for cryomodule
  plasma processing.
  CGC: compressed gas cylinder;
  FP: fore-pump;
  GF: gas filter;
  GV: gate valve;
  HF: HEPA filter;
  MFC: mass flow controller;
  LV: leak valve;
  MV: metering valve;
  P: pressure sensor;
  PRV: pressure regulation valve;
  RGA: residual gas analyzer;
  SR: safety relief valve;
  TMP: turbo-molecular pump;
  VP: viewport.  Cyan: signals recorded by the data acquisition
  system.  Gray: components turned off for plasma processing.
  \label{F:plumb}}
\end{figure}

\begin{figure}
\includegraphics[width=\columnwidth]{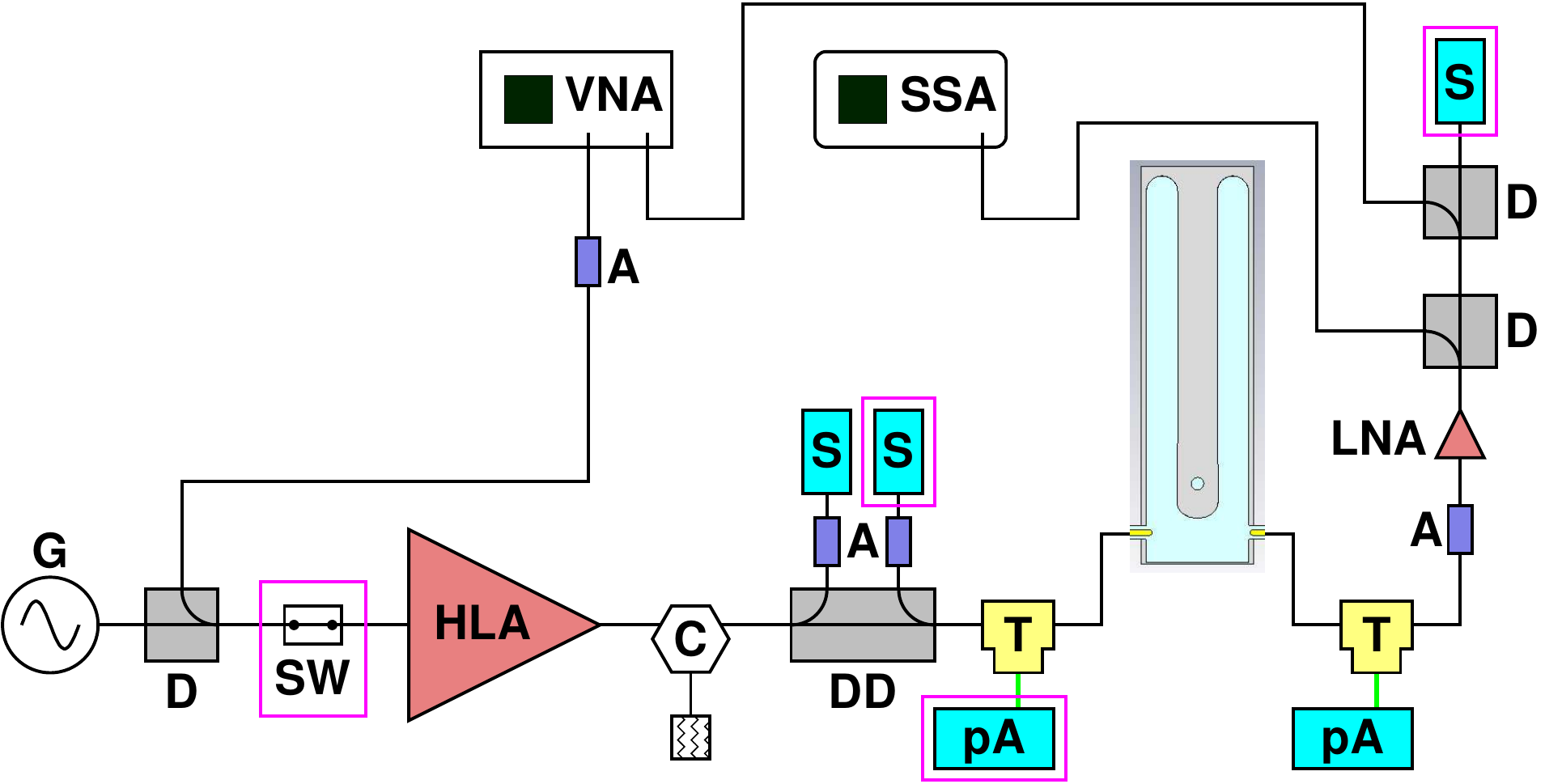}
  
\caption{Schematic of the RF system for cryomodule plasma processing.  A:
  attenuator(s); C: circulator with load; D: directional coupler; DD:
  dual-directional coupler; G: RF signal generator; HLA: high-level
  amplifier; LNA: low-noise signal amplifier; pA: picoammeter; S:
  power sensor; SSA: spectrum analyzer; SW: RF switch; T: bias T; VNA:
  network analyzer.  Signal paths are shown in black (RF) and green
  (dc).  Cyan: signals recorded by the data acquisition system.
  Magenta boxes: components used for the software
  interlock.\label{F:rfsys}}
\end{figure}

\begin{figure*}
\includegraphics[width=\textwidth]{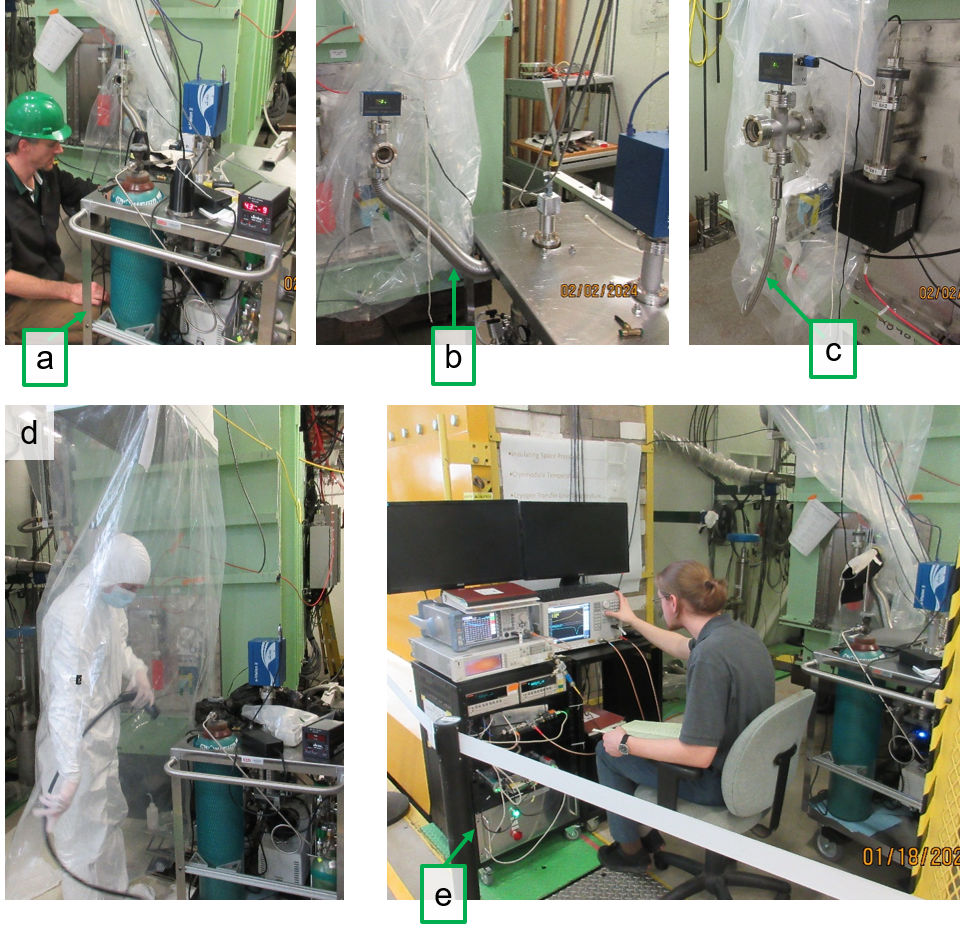}
  
\caption{Cryomodule plasma processing: (a) gas
supply/pumping cart; (b) pumping line connection; (c) gas supply line
connection; (d) clean room hood for beam line connections;
(e) RF cart for plasma generation and monitoring.\label{F:pics}}
\end{figure*}

A new mobile plasma processing system was used to supply and pump the
process gases, drive the plasma with RF power, and monitor the plasma
for the cryomodule.  The circuit for gas supply and pumping is shown
in \cref{F:plumb}; the RF system is shown in \cref{F:rfsys}; photos of
the setup are shown in \cref{F:pics}.

The gas supply and pumping system is similar to the system used in
bench trials \cite{SRF2021:WEPTEV011, NAPAC2022:MOPA91,
  SRF2023:THIXA01}, but the gas flows through a larger volume over
longer distances in the cryomodule case.  Temporary clean room hoods
were used when connecting the gas supply and pumping lines to the
upstream and downstream ends of the cryomodule (\cref{F:pics}d).
Pressure gauges and viewports were added with the gas supply and
pumping connections (\cref{F:pics}b and \cref{F:pics}c).  The
cryomodule insulating space was vented to be sure to avoid cavity
overheating during plasma processing.  As seen in \cref{F:plumb},
inlet gas flow rates were set by mass flow controllers (MFCs).  A
turbomolecular pump (TMP1) was used to pump process gases and
byproducts out of the cavity.  The outgoing gas was sampled by a
residual gas analyzer (RGA), with a leak valve (LV) and second turbo
pump (TMP2) to provide lower pressure for the RGA. To help protect the
cavities from backstreaming of gas in the case of an unplanned
turn-off of the pumps, the gate valves (GV) at the upstream and
downstream ends of the cryomodule were set up to close in case of a
power outage.

The RF system is similar to the version used for on-the-bench plasma
processing.  It includes a 100~W solid state amplifier to drive the
plasma and bias T's to ground the inner conductor of the FPC and
pickup probe and monitor the dc current.  There is a sharp increase in
the FPC current after coupler ignition.  We monitored the resonant
frequency of one or several modes using a network analyzer, following
the approach developed by the Fermilab team
\cite{JAP126:023302}.

We added interlocks to inhibit the RF drive in case of coupler
ignition: (1) a high input coupler current magnitude ($\Ifpc$)
interlock, (2) a low reverse power ($P_r$) interlock, and (3) a low
transmitted power ($P_t$) interlock.  All interlocks detect coupler
ignition; the $P_t$ interlock detects both coupler ignition and loss
of cavity plasma.  The $\Ifpc$ interlock is generally the fastest.
The interlocks were implemented in software with inhibition of the RF
power via an RF switch.  Components used for the interlocks are
highlighted in magenta in \cref{F:rfsys}.

\subsection{Procedure and validation}

\begin{table}

\caption{Process gas parameters.\label{T:recipe}}
\centering
\begin{ruledtabular}
\begin{tabular}{lr}
Inert gas & argon (Ar) \\
Reactive gas & oxygen (O$_2$) \\
Inert gas flow rate & 2.934 mg/min \\
Reactive gas flow rate & 0.326 mg/min \\
Cavity pressure & $\sim 58$ mTorr ($\sim 7.7\tten{-5}$ bar) \\
\end{tabular}
\end{ruledtabular}

\end{table}

\Cref{T:recipe} shows the gas parameters we used for cryomodule plasma
processing.  As seen in \cref{F:plumb}, two gas cylinders supplied the
gas mixture, one with pure Ar and another with a 80\%/20\% Ar/O$_2$
mixture.  The pressure was measured via gauges upstream and downstream
of the cryomodule and near the pump (assuming pure Ar when correcting
for gas composition).

The resulting 90\% Ar/10\% O$_2$ gas mixture at $\sim 60$~mTorr differs from the
95\%
Ne/5\% O$_2$ mixture at $\sim 100$~mTorr used for most of the QWR
plasma development.
Relative to Ne plasma, an Ar plasma allows for a similar plasma
ignition threshold field if the Ar is at lower pressure, seen for our
QWRs as well as HWRs at IMP~\cite{LINAC2016:MOPRC029} and
FRIB~\cite{HIAT2025:TUC03} and elliptical
cavities~\cite{JAP126:023302}.
Lower gas pressure allows for higher plasma
density~\cite{HIAT2025:TUC03}.
Our choice of Ar pressure is lower than used for Ar processing of CEBAF cavities~\cite{SRF2021:TUPTEV004}, but
is within the range studied for PIP-II spoke cavities~\cite{SRF2023:THIXA02}.
The 90\% Ar/10\% O$_2$ mixture is within the range previously used for CEBAF
cavities \cite{SRF2023:WEPWB054}.
Our mass flow rate corresponds to a molar flow rate of about 1.9 sccm,
which is lower than the flow ranges of 5 to 30 sccm used at SNS
\cite{LINAC2016:WE2A03} and Jefferson Lab \cite{SRF2021:TUPTEV004}.  A
lower flow rate may have the advantage of providing a better
signal-to-background ratio for by-products monitored by the RGA\@.
We must emphasize that we have not yet carried out systematic studies
to identify an optimum gas mixture and flow rate for plasma cleaning
of FRIB cavities.

Plasma processing steps were (i) ramp up the RF power and ignite
cavity plasma, (ii) set the RF power to the desired level, (iii) shift
the drive frequency up to increase plasma density, (iv) process with
continuous-wave (CW) RF power for 1 hour.

Commissioning and validation of the new plasma processing system,
procedures, and interlocks were done via bench plasma processing of a
FRIB QWR (S85-982) prior to starting work on the cryomodule.
Before-and-after cold tests showed no field emission X-rays before
plasma processing, and none after plasma processing.

In the cryomodule (SCM813), 5 iterations per cavity were done, using
one or both modes (consecutively), for a total of 5 or 10 hours of
plasma processing per cavity.  
Processing was done on 4 cavities (5-8) with field emission X-rays.
The remaining cavities (1-4) were not treated with plasma, but all
cavities were checked in the before-and-after cold tests, as described
in \cref{S:colds}.

\section{Plasma process measurements and monitoring}

\subsection{RF power and frequency shift\label{S:PfPdFreq}}

\begin{table*}
  \caption{Measured forward power ($P_f$), drive frequency shift
    ($\Delta f_d$), and calculated power dissipation ($P_d$) during
    cryomodule plasma processing; $\Ncit$ = number of trips due to coupler
    ignition.\label{T:pfpd}}

\centering
\begin{ruledtabular}
\begin{tabular}{cccccccccc}
&  \multicolumn{4}{c}{\textbf{404 MHz}} && \multicolumn{4}{c}{\textbf{605 MHz}}\\ \cline{2-5}\cline{7-10}
       & $P_f$     & $\Delta f_d$ & $P_d$           &  &&  $P_f$         & $\Delta f_d$ & $P_d$  & \\
Cavity & (W)       & (MHz)       & (W)             & $\Ncit$ && (W)            & (MHz)       & (W) & $\Ncit$ \\ \hline
8       & 12.0 $\pm$ 1.2 & 1.0 $\rightarrow$ 0.8 & 0.15 $\pm$ 0.012 &6&& 10.8 $\pm$ 0.5 & 4.2 $\rightarrow$ 3.5  & 0.53 $\pm$ 0.05 &0\\
7       & 11.9 $\pm$ 1.2 & 0.8                   & 0.14 $\pm$ 0.011 &2&&    -           &-&  -              &-\\
6       & -              &  -                    &   -              &-&& 10.9 $\pm$ 0.5 & 3.9 $\rightarrow$ 3.8 & 0.55 $\pm$ 0.06 &0\\
5 init  & 12.7 $\pm$ 1.3 & 0.9 $\rightarrow$ 0.8 & 0.16 $\pm$ 0.013 &4&&
                                                                       15.4 $\pm$ 0.7 & 3.6                   & 0.65 $\pm$ 0.07 &0\\
5 fin   & 12.2 $\pm$ 1.2 & 0.8                   & 0.17 $\pm$ 0.013 &1&& 22.5 $\pm$ 1.0 & 4.9 $\rightarrow$ 5.1  & 0.83 $\pm$ 0.08 &1\\
\end{tabular}
\end{ruledtabular}
\end{table*}

Summary information about the forward power, drive frequency shift,
and power dissipation during plasma processing is provided in
\cref{T:pfpd}\@, along with statistics on trips due to coupler
ignition.  We processed near the maximum frequency shift
($\sim 0.9$~MHz for the 404~MHz mode and $\sim 3.5$~MHz for the
605~MHz mode) in order to maximize the plasma density.  Operating near
the frequency limit, the drive frequency is approximately equal to the
shifted resonant frequency.

The plasma density can be inferred from
the resonant frequency shift, which is a result of the decrease in the
effective permittivity due to the plasma~\cite{LIEBLICHT2005:PrincPlasDisMatProc,
  SRF2013:TUP057}.  In the case of a uniform plasma density,
the effective permittivity $\epsilon$ is related to the unperturbed
resonant frequency
$f_0$ and perturbed resonant frequency $f$ via
\begin{equation}
\frac{\epsilon}{\epsilon_0} = \left(\frac{f_0}{f}\right)^2\, ,
\end{equation}
where $\epsilon_0$ is the permittivity of free space.
Making use of the relationship between $\epsilon$ and the 
electron number density $n_e$, we can express the latter in terms of 
$f_0$ and $f$ via
\begin{equation}
n_0 = 4 \pi^2 \frac{\epsilon_0 m_e}{q_e^2} \left(f^2 - f_0^2\right)\, ,
\end{equation}
where $m_e$ is the electron mass and $q_e$ is the electron charge.
We calculate 
$n_e \approx 0.9\tten{13}$~m$^{-3}$ for the 404~MHz mode and $n_e \approx 5\tten{13}$~m$^{-3}$
for the 605~MHz mode.  As the plasma distribution is not uniform,
these densities can be considered to be estimates, weighted by the
field distribution of the drive mode.

The error estimates for the forward and dissipated power in
\cref{T:pfpd} are based on the estimated systematic errors in the RF
power measurements (\cref{A:errs}).  The power dissipation was
calculated indirectly, as discussed in \cref{A:Pd}.

The $P_f$ and $P_d$ values in \cref{T:pfpd} are approximate averages
over a total of 5 hours of processing.  An exception is Cavity 5,
which we processed with lower $P_f$ and $\Delta f_d$ in the initial
rounds (1-2) and higher $P_f$ and $\Delta f_d$ in the final rounds
(3-5); two rows are accordingly provided for Cavity 5 in
\cref{T:pfpd}.  Plots of $P_f$ and $P_d$ as a function of time are
included in \cref{A:PowPick}.  With steady plasma, $P_f$ is relatively
stable.  We observe some increase in $P_d$ over time, correlated with
a more pronounced increase in the RF stored energy ($U$, inferred from
the transmitted power $P_t$).  The increase in $U$ is more noticeable
for the 404~MHz mode, being of order 2~dB.

The ranges given in \cref{T:pfpd} for the drive frequency shifts
indicate that we were generally able to process at higher $\Delta f_d$
early on and lower $\Delta f_d$ later, with Cavity 5 being an
exception.

The values of $P_d$ in \cref{T:pfpd} include both the power
dissipation in the walls of the cavity and the power dissipation by
the plasma.  We estimate that the power dissipation in the plasma is
about 0.75 $P_d$ for the 404~MHz mode and about 0.90 $P_d$ for the
605~MHz mode.  The basis for these estimates is discussed in 
\cref{A:Px}.  Plots of estimated power dissipation in the plasma
($P_x$) as a function of time are included in \cref{A:PowPick}.
Generally $P_x$ tracks $P_d$ closely, though this may be a result of
the assumptions made in the calculations.

\begin{figure}
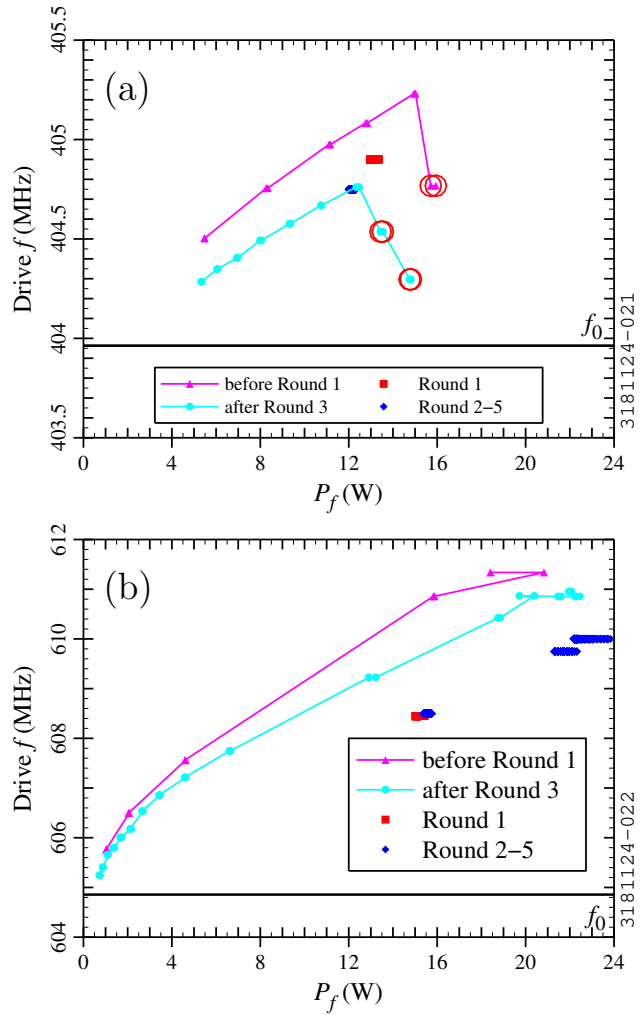

\GRAFlabelcoord{15}{255}
\GRAFoffset{-65}{-65}
\GRAFwidth[\landwidth]{490}{390}
\incGRAFlabel{c5_404_th_f_proc_fdpf_fdc_lh_a}{(a)}\\[-2ex]
\incGRAFlabel{c5_605_th_f_proc_fdpf_fdc_lh_a}{(b)}\\[-1.5ex]
  
\caption{Forward power and drive frequency for Cavity 5 using (a) the
  404~MHz TEM mode and (b) the 605~MHz dipole mode.  Light markers:
  maximum frequency shift with cavity plasma.  Red circles: coupler
  ignition.  Dark markers: drive frequency and forward power during
  plasma processing.  Black lines: unshifted resonant frequency.  Note
  the different vertical scales for the 2 modes.\label{F:flim}}
\end{figure}

We measured the ``frequency limits'' for each cavity before the first
and after the third round of processing.  Some examples are shown in
\cref{F:flim}.  In these measurements, we ramped up the RF power while
driving the mode on resonance (the black lines indicate the unshifted
resonant frequency $f_0$), ignited cavity plasma, adjusted the forward
power, ramped up the drive frequency to measure the maximum frequency
shift before loss of cavity plasma, and repeated the steps for
different $P_f$ values.  As seen in \cref{F:flim}, we are generally
able to achieve a higher frequency shift (and hence a higher plasma
density) with higher RF power up to the coupler ignition threshold.
(Coupler ignition trips seen during the dedicated frequency limit
measurements are not included in the trip statistics of
\cref{T:pfpd}.)

For the 404~MHz mode, we transitioned to coupler plasma at high
forward power (\cref{F:flim}a, red circles); otherwise we
transitioned from cavity plasma back to neutral gas at the frequency
limit.  Thus coupler ignition limits the plasma density, with the
highest plasma density obtained for $P_f \sim 11$ to 12~W.

For the 605~MHz mode, the highest plasma density is obtained with the
highest forward power (for our system, 605~MHz was above the design
frequency range of the RF amplifier and circulator, such that the
available power was less than for the 404~MHz case).

Initially, we were able to reach higher frequency shift
(\cref{F:flim}, magenta triangles), but we found that the plasma
was not stable enough for 1-hour processing sessions.  After 3
iterations of plasma processing, we measured lower frequency limits
(\cref{F:flim}, cyan circles).  We observed some fluctuations in
the cavity and coupler ignition thresholds between different cavities
and different iterations.

The red squares and blue diamonds in \cref{F:flim} indicate the
drive frequency and power used for plasma processing.  The horizontal
spread is due to the tendency for the RF power to drift downward
during 1 hour of plasma processing.  We tried to find a balance
between processing at the highest possible frequency shift versus
processing with lower frequency shift for the sake of better plasma
stability and reduced risk of plasma extinction or coupler ignition.
We had a tendency to err on the side of optimism, and therefore often
had to back off on the frequency shift during plasma iterations, as
seen in \cref{F:flim}a.  The dipole mode in Cavity 5
(\cref{F:flim}b) was an exception in which we initially erred on
the side of pessimism and became more optimistic for later iterations;
we were able to process with higher power and higher frequency shift
while maintaining a comfortable margin between the drive frequency
shift and the maximum frequency shift.

\subsection{Monitoring of reaction by-products}

\begin{figure}
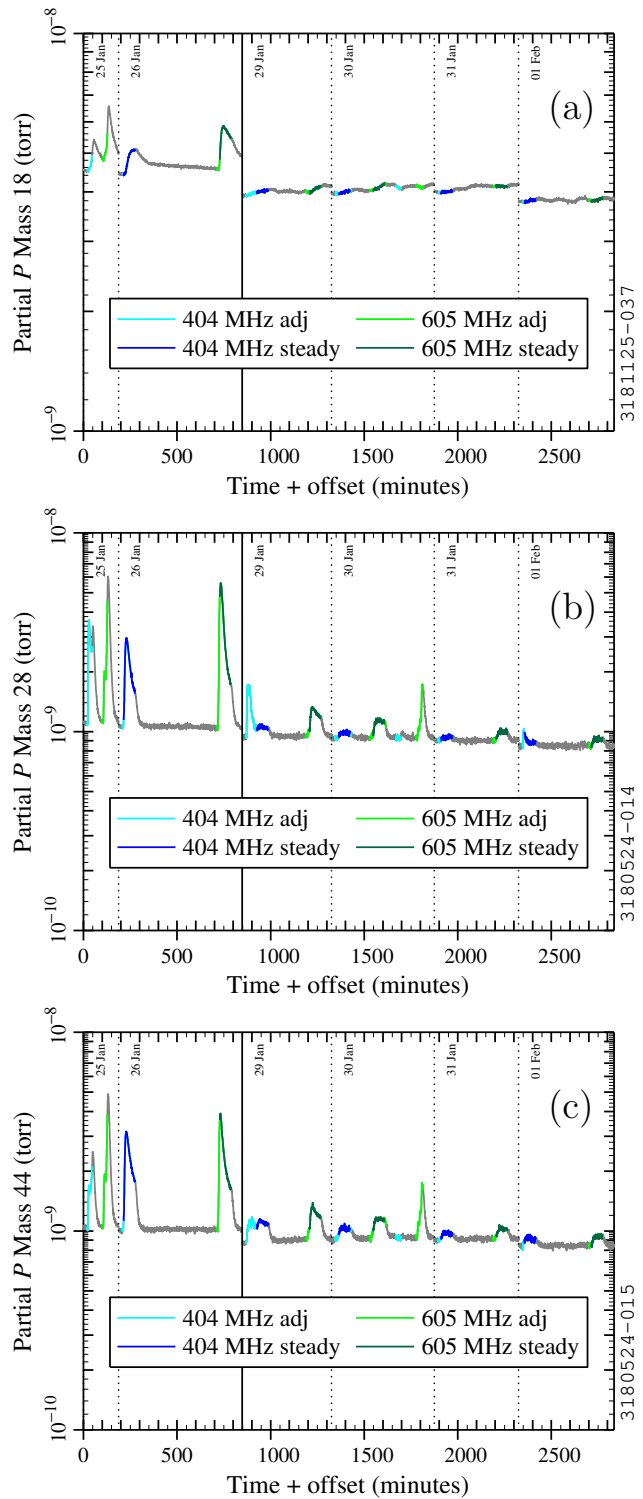

\GRAFlabelcoord{350}{235}
\GRAFoffset{-65}{-65}
\GRAFwidth[\landwidth]{490}{390}
\incGRAFlabel{clean_cav5_m18a}{(a)}\\[-2ex]
\incGRAFlabel{clean_cav5_m28a}{(b)}\\[-2ex]
\incGRAFlabel{clean_cav5_m44a}{(c)}\\[-1.5ex]
  
\caption{RGA signals for (a) Mass 18 (H$_2$O), (b) Mass 28 (CO, N$_2$) and (c) Mass 44
  (CO$_2$) during plasma processing of Cavity 5.  Dark blue, dark
  green: steady plasma; light blue, light green: adjusting drive power
  and drive frequency; gray: plasma off.  The time axis is adjusted to
  zoom in on intervals with plasma on (vertical lines indicate
  temporal ``cuts''). \label{F:rgaC5}}

\end{figure}

As seen in \cref{F:plumb}, we used a residual gas analyzer (RGA) to
sample the gas pumped out of the cavities.  The pressure in the RGA
chamber was about $1.8 \cdot 10^{-6}$~Torr ($2.4 \cdot 10^{-9}$~bar).
As shown in \cref{F:rgaC5}, clear peaks in the RGA signals for Mass 44
(CO$_2$) and Mass 28 (CO or N$_2$) were seen with plasma on; the peaks
decreased with time and with iterations.  The peaks for Mass 28 and 44
tended to drop to lower values when we used both the 404~MHz and the
605~MHz modes on same cavity.  Smaller peaks for Mass 18 (H$_2$O) were
seen in the first iteration, but there was little or no signal in
subsequent iterations.

\Cref{F:rgaC5} shows that peaks in Mass 28 and Mass 44 were produced
not only with steady plasma, but also when we adjusted the RF power
and frequency shift for frequency limit measurements prior to the
first iteration and after the third iteration.  The RGA response tended
to be larger for the 605~MHz mode, even though processing was done
with the 404~MHz mode first.

\begin{figure}
\includegraphics[width=\columnwidth]{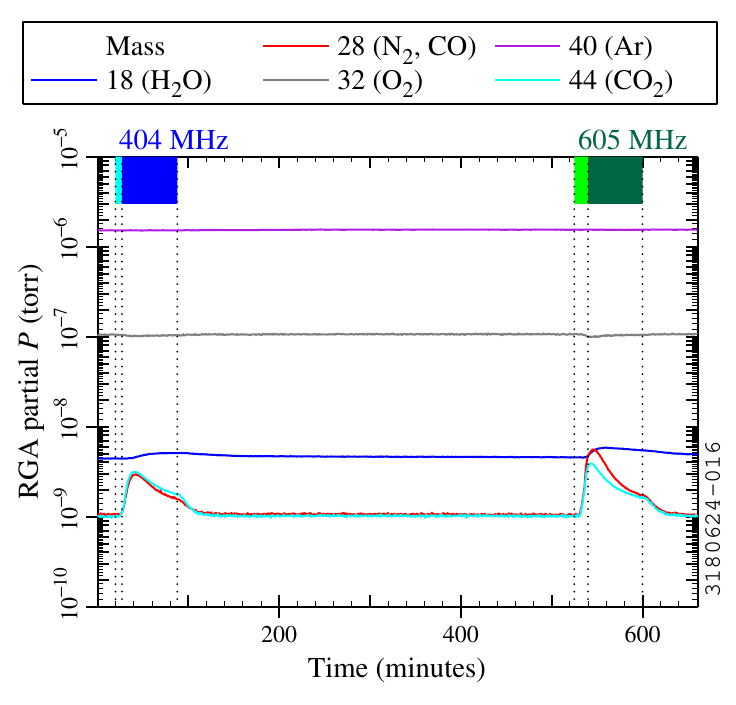}
  
\caption{Zoomed-in view of RGA signals for selected masses in the
  first round of plasma processing for Cavity 5.  The shaded areas and
  dotted lines indicate RF power and frequency ramp-ups (lighter
  shades) and steady plasma (darker shades).\label{F:rgaC5:1}}
\end{figure}

\Cref{F:rgaC5:1} shows the RGA signals for several different masses
in the first round of plasma processing on Cavity 5.  Mass 28 and Mass
44 show the largest response.  A small increase in Mass 18 (H$_2$O)
can be seen, with a slower response time (consistent with \cref{F:rgaC5}a).  The slower response may
indicate that heating of the cavity walls by the plasma contributes to
water production.  A small decrease in Mass 32 (O$_2$) can be seen as
well, evidently due to some of the oxygen being consumed as CO$_2$ and
CO are produced.  In contrast, the signal for Mass 40 (Ar) is steady.
Though changes are clearly visible in the first round, Mass 18 and
Mass 32 showed little or no response to the plasma in subsequent
rounds.

\subsection{Plasma monitoring: light and current\label{S:lightI}}

\begin{figure}
\GRAFlabelcoord{10}{160}
\GRAFwidth[0.875\columnwidth]{213}{200}
\incGRAFlabel{img605_c8_r3crop}{\textcolor{white}{(a)}}\\[1ex]
\incGRAFlabel{img605_c8_r4crop}{\textcolor{white}{(b)}}\\[1ex]
\incGRAFlabel{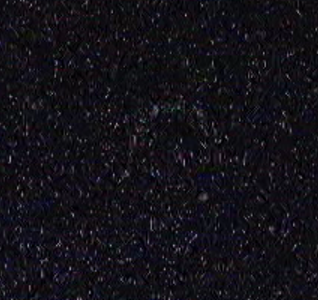}{\textcolor{white}{(c)}}\\[1ex]
  
\caption{Images of the plasma in Cavity 8 driven by the 605~MHz mode
  in (a) Round 3 and (b) Round 4 or (c) by the 404~MHz mode in Round 3.\label{F:light}}
\end{figure}

We monitored the light from the plasma using viewports on the ends of
the cryomodule (as seen in \cref{F:plumb}); this was easiest for
Cavity 8, which was the closest to the upstream end.  Two images are
shown for the 605~MHz case in \cref{F:light}a and \cref{F:light}b.  There is a clear
left-right asymmetry in the light, from which we can infer that the
plasma is likely present in the left or right lobe of the dipole mode
(per \cref{F:maps}b, front view), but not both.  Similar behavior for
the 605~MHz case could be seen in the bench trials.  In the 404~MHz
case (\cref{F:light}c), the light was rather dim, but was more left-right symmetric,
consistent with the field distribution (\cref{F:maps}a).

\begin{figure}
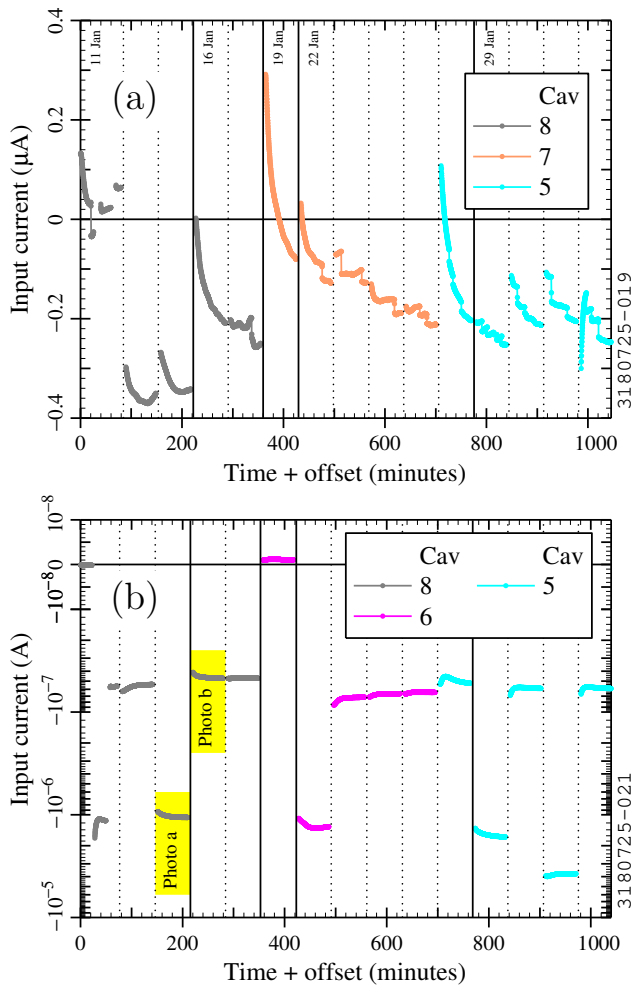

\GRAFlabelcoord{15}{225}
\GRAFoffset{-65}{-65}
\GRAFwidth[\landwidth]{490}{390}
\GRAFlabelbox{40}{30}%
\incGRAFboxlabel{cc_404_rf_bin_fdcor_it_mixc_in}{(a)}\\[-2ex]
\incGRAFboxlabel{cc_605_rf_bin_fdcor_it_shad_in}{(b)}\\[-1.5ex]

\caption{Measured dc current collection by the FPC as a function of time with
steady plasma for (a) the 404 MHz mode and (b) the 605 MHz mode.\label{F:Ifpc}}
\end{figure}

As seen in \cref{F:rfsys}, we could monitor the dc current from the
FPC and pickup antenna via bias T's and picoammeters.  The measured
current from the FPC with steady plasma is shown in \cref{F:Ifpc}; see
\cref{A:PowPick} for the pickup current.  Time cuts are used to zoom
in on the steady-plasma intervals; the dotted vertical lines indicate
time cuts of about 1 day and the solid vertical lines denote longer
time cuts.  (Note that the time offsets are different between steady
plasma for 404~MHz and steady plasma for 605~MHz, with the former
preceding the latter.)

For the 404~MHz mode (\cref{F:Ifpc}a), $\Ifpc$ started out positive
and slowly drifted down to a steady-state value of order $-200$~nA,
though not always returning to the same value between the end of one
plasma iteration and the start of the next iteration.  We can
speculate that the downward drift in $\Ifpc$ could be due to reduction
in the secondary electron yield of the copper FPC antenna.

In the 605~MHz case (\cref{F:Ifpc}b), the FPC current tended to be of
order $-1$~$\mu$A, of order $-100$~nA, or of order 1~nA in magnitude.
The current was correlated with the light distribution.  For example,
it was high in magnitude when reflected light was seen on the left
(\cref{F:light}a) and lower in magnitude when reflected light was seen
on the right (\cref{F:light}b); the yellow highlights in
\cref{F:Ifpc}b indicate the intervals during which the still images of
\cref{F:light} were recorded.  As the FPC is on the right side
relative to the photos, this makes sense---reflected light opposite
the FPC due to plasma on the FPC side of the cavity is correlated with
an FPC current that is larger in magnitude.  As seen in \cref{F:Ifpc},
there is some drift in the FPC current over time for the 605~MHz mode,
but the effect is less pronounced than that seen for the 404~MHz mode.

The dc current from the pickup antenna (\cref{A:PowPick}) is much
smaller in magnitude than the FPC current, consistent with the smaller
cross-sectional area and length of the pickup probe relative to the
FPC antenna.  For the 404~MHz mode, the pickup current is positive and
tends to drift up over time toward a value of order 6~nA.  The drift
direction is opposite that of $\Ifpc$, but the currents do not mirror
one another closely.  We can again speculate that the drift is
associated with a change in the secondary electron yield of the probe
antenna (stainless steel).  In the 605~MHz case, the pickup current
tends to show more drift over time than the FPC current, sometimes
changing from negative to positive.  As is the case for the 404~MHz
mode, the drift direction is generally upward.  We would expect the
FPC and pickup current magnitudes to follow opposite trends if plasma
is forming on the FPC side in some cases and on the pickup side on
other cases, but this cannot be seen very clearly in the measurements;
the drift in the pickup current (as well as changes in sign) make the
observations more difficult to interpret.

Our diagnostics support the interpretation that, for the 404~MHz TEM mode, we always
ignite the bottom portion of the cavity, where the surface electric
field is highest.  For the 605~MHz dipole mode, there is some
randomness as to which portion of the cavity ignites, as exemplified
by \cref{F:light} and \cref{F:Ifpc}.  The cases with very low FPC
current likely correspond to plasma distributions in the upper half of
the cavity, far from the FPC\@.  When measuring the dipole mode
frequency limits (\cref{S:PfPdFreq}), we did not see any change in
the maximum drive frequency shift for different the plasma locations
(though likely we would if the frequency shift was limited by coupler
ignition),

\subsection{Plasma ignition threshold measurements\label{S:th}}

\begin{table*}
\caption{Measured cavity ignition thresholds during cryomodule plasma
  processing.  $P_f^*$ = forward power for cavity ignition; $U^*$ =
  stored energy at cavity ignition inferred from $P_t$.  The measured FPC
  coupling strength ($\Qext{1}$) is included for reference.\label{T:cavi}}
\centering
\begin{ruledtabular}
\begin{tabular}{cccccccc}
&  \multicolumn{3}{c}{\textbf{404 MHz}} && \multicolumn{3}{c}{\textbf{605 MHz}}\\ \cline{2-4}\cline{6-8}
       & $P_f^*$ & $U^*$      &           && $P_f^*$ & $U^*$      & \\
Cavity & (W)   & ($\mu$J) & $\Qext{1}$ && (W)   & ($\mu$J) & $\Qext{1}$ \\ \hline

8 & $23.0 \pm 1.8$ & $3.01 \pm 0.25$ & $6.08\tten{5}$ && $5.04 \pm 0.37$ & $5.38 \pm 0.39$ & $9.91\tten{4}$ \\
7 & $23.1 \pm 1.9$ & $2.75 \pm 0.21$ & $6.28\tten{5}$ &&       -         &       -         &      -         \\
6 &      -         &      -          &      -         && $4.65 \pm 0.41$ & $5.42 \pm 0.50$ & $9.00\tten{4}$ \\
5 & $23.8 \pm 2.5$ & $3.11 \pm 0.32$ & $5.66\tten{5}$ && $4.84 \pm 0.39$ & $5.36 \pm 0.44$ & $9.07\tten{4}$ \\
\end{tabular}
\end{ruledtabular}
\end{table*}

Plasma processing iterations and frequency limit measurements allowed
us to measure the cavity plasma ignition threshold repeatedly for each
cavity and drive mode.  Some statistics are included in \cref{T:cavi}.
The values of forward power at ignition ($P_f^*$) and cavity stored
energy at ignition ($U^*$) are averages over $>25$ measurements.  The
(sample) standard deviation is included for both.  Systematic errors
are not included in \cref{T:cavi}, but the systematic error in $P_f^*$
is estimated to be $\pm 10$\% for 404~MHz and $\pm 4$\% for 605~MHz;
the systematic error in $U^*$ is estimated to be $\pm 12$\% for
404~MHz and $\pm 20$\% for 605~MHz (additional information on
systematic error estimates is included in \cref{A:errs}).

We expect the same $U^*$ for each cavity, and find that, indeed, the
measured values are within the statistical errors.  More spread is
seen in the 404~MHz mode, which may be due to cavity-to-cavity
differences: the 404~MHz mode may be more sensitive to the tuning
plate position than the 605~MHz mode, given that the highest electric
field for the 404~MHz mode is on the inner conductor, opposite the
tuning plate (\cref{F:maps}a), in contrast to the 605~MHz mode, for
which the highest field regions are further up along the inner
conductor (\cref{F:maps}b).

We would expect $P_f^*$ to be correlated with $\Qext{1}$.
This can be seen to some extent for the 605 MHz mode, but not for the
404 MHz mode.  The lack of correlation for the latter may be, again,
due to cavity-to-cavity differences.

Based on the values of $U^*$ in \cref{T:cavi} and the CST model, we
estimate a peak surface electric field at ignition of 14 kV/m for the
404 MHz mode and 18.5 kV/m for the 605 MHz mode.

In some cases, we observed a systematic increase in $P_f^*$ and $U^*$
during the first 2 plasma processing iterations (as much as 10 to 15\%
in $P_f^*$).  In other cases, there was no obvious trend in time
relative to the scatter in the measured ignition threshold.

\section{Before-and-after cold tests\label{S:colds}}

Cold testing of the cryomodule was done in the test bunker before and
after cryomodule plasma processing.  The cryomodule was installed in
the FRIB driver linac in the Summer 2024 maintenance period in order
to remove and refurbish one of the original QWR cryomodules.  In-tunnel
cold testing of the cryomodule was done in September 2024 prior to
resumption of user operations.

\subsection{Field emission}

\begin{figure}
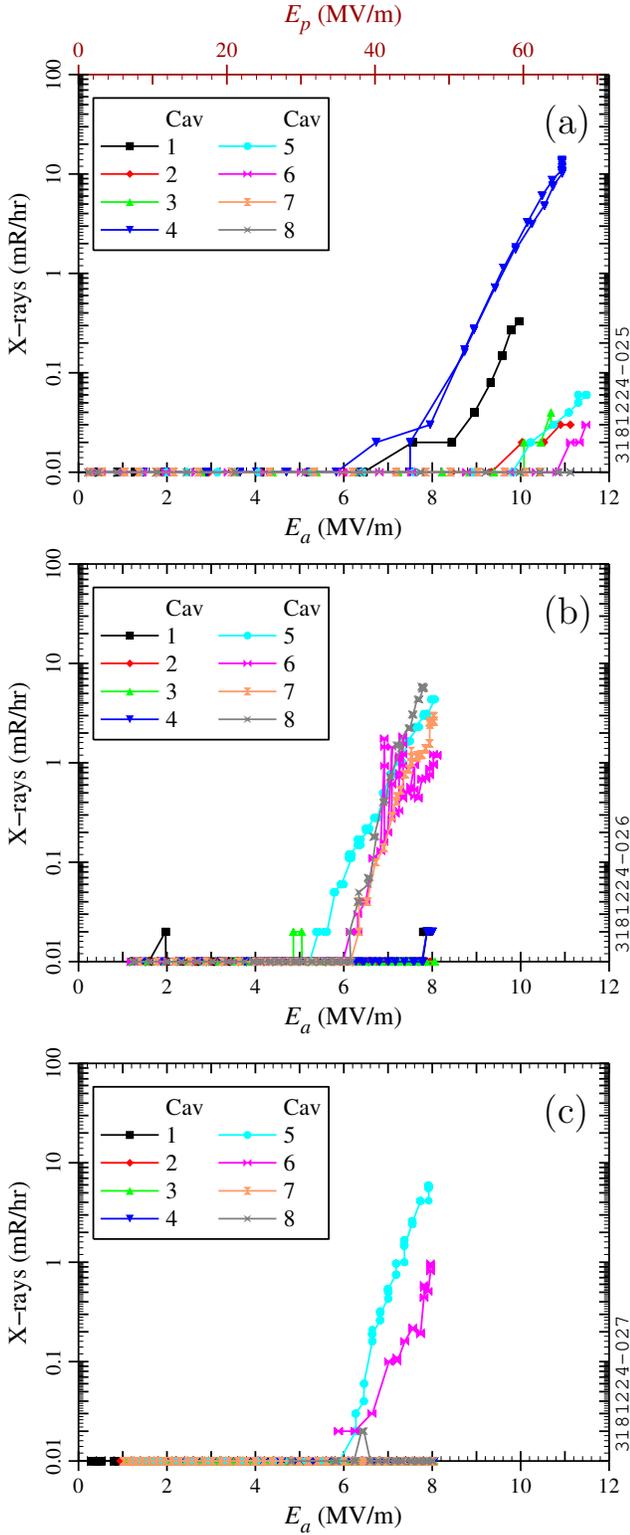

\GRAFlabelcoord{350}{255}
\GRAFoffset{-65}{-58}
\GRAFwidth[\triwidth]{490}{415}
\incGRAFlabel{b085_813_e20_xebb}{(a)}\\[-2ex]
\GRAFoffset{-65}{-65}
\GRAFwidth[\triwidth]{490}{390}
\incGRAFlabel{cw_aut_cm813t2f_xe_b}{(b)}\\[-2ex]
\GRAFoffset{-65}{-65}
\GRAFwidth[\triwidth]{490}{390}
\incGRAFlabel{cw_aut_cm813t3f_xe_b}{(c)}\\[-1.5ex]
  
\caption{Measured X-rays as a function of RF field in cold tests: (a)
  cavity certification tests and bunker tests (b) before and (c) after
  plasma processing.  The background X-ray level is 0.01~mR/hour, with
  signals above $\sim 0.02$~mR/hour being significant.  $E_a$ =
  accelerating gradient; $E_p$ = peak surface electric
  field.\label{F:colds}}
\end{figure}

\Cref{F:colds}a shows X-ray measurements in the final cavity
certification test for each of the cavities before installation in the
cryomodule.  The cavities showed no field emission X-rays below
$E_a = 7$ to 8~MV/m.  \Cref{F:colds}b shows corresponding measurements
in the cryomodule bunker tests before plasma processing.  Cavities 5
through 8 show X-ray onset fields of 5 to 6~MV/m, likely due to some
contamination introduced between the cavity tests and the bunker test.
(High-risk in-clean-room steps between cavity testing and bunker
testing include venting of the cavity, installation of the FPC, and
assembly of the cavity onto the cold mass.)

\Cref{F:colds}c shows the bunker measurements after plasma
processing of Cavities 5 through 8.  \Cref{T:proc} provides summary
information about plasma processing and field emission onsets.  The
unprocessed cavities (1-4) showed background X-rays before and after
(up to the maximum field of 8~MV/m).  Cavity 7 (orange hourglasses)
and Cavity 8 (gray crosses) improved significantly after processing;
Cavity 5 (cyan circles) showed a small improvement; Cavity 6 (magenta
bow-ties) showed similar performance before and after processing.  The
X-ray measurements thus indicate that field emission was reduced with
plasma processing for 3 out of 4 cavities.

\begin{table}
  \caption{Cryomodule plasma processing modes, durations, and field
    emission onsets in cryomodule cold tests.  $E_a$ = accelerating
    gradient.\label{T:proc}}

\centering
\begin{ruledtabular}
\begin{tabular}{ccccc}
 &  & Total  & \multicolumn{2}{c}{FE onset $E_a$}\\
 & Mode(s) & process & \multicolumn{2}{c}{(MV/m)}\\ \cline{4-5}
Cavity & (MHz) & time (hrs) & before & after\\ \hline
8 & both & $\sim 10$ & $\sim 6.0$ & $\geq 8.0$ \\
7 & 404 & $\sim 5$ & $\sim 6.2$ & $\geq 8.0$ \\
6 & 605 & $\sim 5$ & $\sim 6.0$ & $\sim 6.0$ \\
5 & both & $\sim 10$ & $\sim 5.3$ & $\sim 6.0$ \\
\end{tabular}
\end{ruledtabular}

\end{table}

X-ray measurements showed no performance degradation between the
post-plasma bunker test and the tunnel test (the latter was done up to
$E_a = 6.7$~MV/m).  Cavity 5 may have improved a bit, as it showed no
X-rays up to 6.7 MV/m in the tunnel test.  The field emission onset
for Cavity 6 was similar for the last bunker test and the tunnel test.

We infer from the cold tests that processing with only the 605~MHz
mode (as was done for Cavity 6) may be less effective for field
emission mitigation.  However, it is difficult to draw strong
conclusions based on four cavities; we note that Cavity 5 and Cavity 8
had the same treatment with different results.

\subsection{Multipacting}

The $\beta_m = 0.086$ QWR has 3 multipacting barriers.  Generally we are
able to jump over the low barrier (at $E_a \sim 5.5$~kV/m) without
conditioning it.  The middle barrier ($E_a \sim 70$~kV/m) and high
barrier ($E_a \sim 0.75$~MV/m) typically require of order 1 hour to
condition during cavity certification tests, though the conditioning
time can vary significantly from one cavity to another
\cite{NIMA1014:165675}.  Conditioning can be done more efficiently in
the cryomodule due to the stronger coupling of the FPC relative to the
input couplers used in Dewar tests.

\begin{figure}
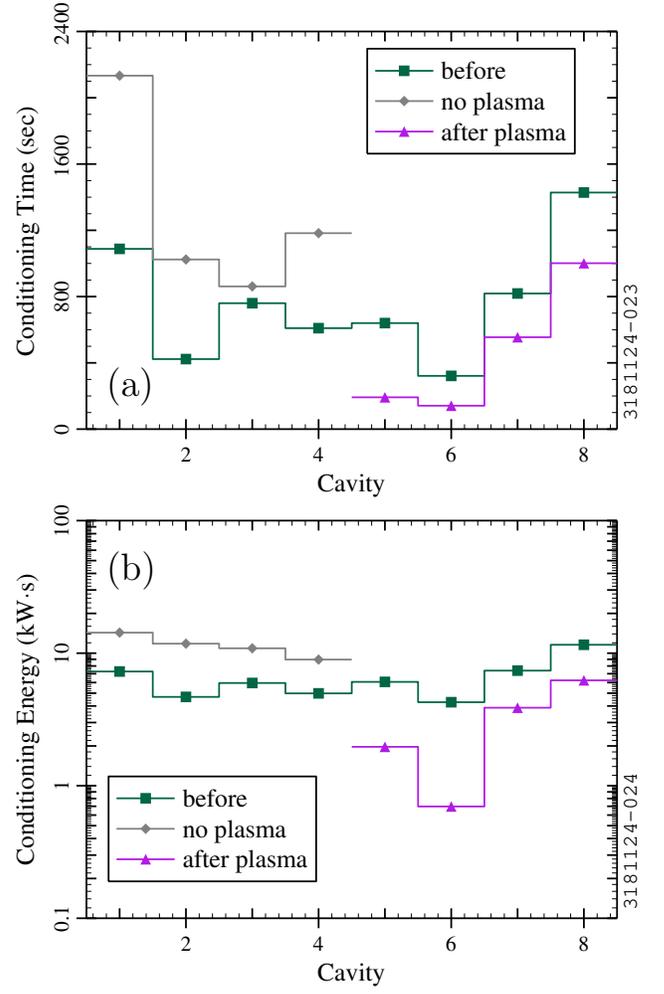

\GRAFoffset{-65}{-65}
\GRAFwidth[\landwidth]{490}{390}
\GRAFlabelcoord{15}{25}
\incGRAFlabel{ylc_hb_t_pd_scm813_bar_lh_p1}{(a)}\\[-3ex]
\GRAFlabelcoord{15}{255}
\incGRAFlabel{ylc_hb_t_pd_scm813_bar_lh_p2}{(b)}\\[-2ex]
  
\caption{High multipacting barrier conditioning statistics from
  cryomodule cold tests: (a) time to condition and (b) time integral
  of power dissipation into multipacting.  Green squares: before
  plasma processing.  Purple triangles, gray diamonds: after plasma
  processing of Cavity 5 through 8.\label{F:mp:high}}
\end{figure}

\Cref{F:mp:high}a shows the time needed to condition the high
barrier in the bunker test before plasma processing (dark green
squares) and in the tunnel test after plasma processing (gray
diamonds, purple triangles).  After plasma processing, Cavities 5
through 8 could be conditioned more rapidly (purple triangles), in
contrast to Cavities 1 through 4, which were not plasma processed and
took longer to condition (gray diamonds).

As the high barrier can generally be conditioned more rapidly with
higher RF power, the conditioning time alone is not the best indicator
of barrier strength.  For more of an Apples-To-Apples comparison,
\cref{F:mp:high}b shows the integral over time of the estimated power
dissipated in the high barrier during conditioning.  The power
dissipation into multipacting is typically of order 1 to 10~W during
conditioning with the FPC, tending to decrease over time with constant
forward power as the barrier weakens.  The integral of conditioning
power shows consistent trends: less conditioning energy is needed
after plasma processing (purple triangles), in contrast to the
unprocessed cavities (gray diamonds).

Thus, there is a clear weakening of the high barrier due to plasma
processing.  Interestingly, the easiest conditioning case (in terms of
both conditioning time and conditioning energy) is Cavity 6, which was
plasma processed with the 605 MHz dipole mode only, with no significant
reduction in field emission.

\Cref{F:mp:high} indicates that the conditioning time was longer
after plasma processing for the cavities which did not receive plasma
processing.  This could be due to migration of H$_2$O from the
processed cavities to the unprocessed cavities.  As seen in
\cref{F:plumb}, the gas flow was oriented such that the reaction
by-products were pumped through the unprocessed cavities.

\section{Conclusion}

A step-wise effort to develop in-situ plasma processing capability for
the FRIB superconducting linac has been undertaken.  Results so far
suggest that plasma processing has good potential for improving the
performance of FRIB cryomodules.  A first plasma processing trial was
done on a spare FRIB quarter-wave resonator cryomodule using two
higher-order modes to drive the plasma.  Before-and-after cold tests
showed a significant reduction in field emission after plasma
processing, along with a significant reduction in the time needed to
condition the high multipacting barrier.

The TEM $5\lambda/4$ ($\sim 404$~MHz) mode appears well-suited for
reduction of field emission via plasma processing; a dipole mode at
$\sim 605$~MHz appears helpful for multipacting but less useful to
reduce field emission.  We are able to reach higher plasma density and
higher power transfer into the plasma with the dipole mode, but the
plasma distribution appears less favorable for cleaning surfaces that
are likely to produce field emission.

We are investigating other HOMs and ``dual-tone'' plasma generation
\cite{LINAC2024:TUPB011} for their potential to provide more effective
processing.  We plan to do a first plasma processing trial in the FRIB
tunnel in the next long maintenance period.  Plasma processing shows
good potential to save significant down time and refurbishment labor
if performance degradation occurs during long-term FRIB linac
operation.

\begin{acknowledgments}

Early FRIB plasma development, measurements, and analysis were lead
by Cong Zhang.  John Popielarski provided valuable guidance and
support for the FRIB plasma efforts until his untimely passing in
2022.  Jacob Brown and Sara Zeidan assisted with plasma measurements
and development.  We thank Pete Donald, James Erdelean, Dave Norton,
John Schwartz, and Dean Thelan for their support and assistance.
Igor Nesterenko provided valuable help and guidance with digital
cameras and optics for plasma imaging and monitoring.  Dan Morris
and John Brandon provided expertise and guidance with RF systems.
Our work is a collaborative effort with the FRIB cryogenics team,
the FRIB cavity preparation team, and the rest of the FRIB
laboratory.

We thank the plasma teams at Oak Ridge National Laboratory, Jefferson
Laboratory, Fermilab, IJCLab, Argonne National Laboratory, and
Brookhaven National Laboratory for useful discussions, information
sharing, and suggestions.  We are especially thankful to colleagues at
Oak Ridge and Fermilab who shared their plasma processing
expertise. The encouragement from Fermilab/BNL colleagues to implement
network analyzer monitoring of the resonant frequency, explore
additional higher-order modes, and consider lower gas pressures was
particularly valuable for our development work.

This material is based upon work supported by the U.S. Department of
Energy, Office of Science, Office of Nuclear Physics, and used
resources of the Facility for Rare Isotope Beams (FRIB) Operations,
which is a DOE Office of Science User Facility under Cooperative
Agreement DE-SC0023633.  Additional support was provided by the State
of Michigan and Michigan State University.
\end{acknowledgments}

\appendix

\section{Error analysis: RF power measurements\label{A:errs}}

An analysis was done to estimate the systematic errors in the measured
RF power values ($P_f, P_r, P_t$).  The results were used to propagate
the errors to the calculated values of stored energy ($U$) and power
dissipation ($P_d$).  The uncertainties in $P_f$ and $P_d$ in
\cref{T:pfpd} are based on the analysis described in this appendix.
The systematic error values given in \cref{S:th} for the ignition
thresholds ($P_f^*$, $U^*$) are based on the same analysis.

\subsection{Uncertainty evaluation: \texorpdfstring{$P_f$}{Pf} and \texorpdfstring{$P_r$}{Pr}}

Directivity errors and mismatch are the major sources of systematic
error for our CW measurements of $P_f$ and $P_r$.  These sources were
considered in recent analyses for measurements on superconducting
cavities by J. Holzbauer and colleagues \cite{NIMA830:22to29,
  NIM913:07to14}.  In our case, we do not consider the mismatch from
imperfect RF components to have a major impact on the measured values
of $\Qext{1}$, $\Qext{2}$, and the low-field $Q_0$, as we are able
to obtain them from network analyzer measurements without additional
components such as circulators and bias T's in the circuit.  Hence we
are concerned primarily with the dual directional coupler directivity
and mismatches between the dual directional coupler and the cavity due
to the adverse impact of the latter on the overall directivity of the
measurements.

As discussed in Ref.~\cite{NIM913:07to14}, a vector method can in
principle be used to correct for directivity; however we are not able
to apply such a method, as we are not equipped to measure RF phases.
Likewise, we do not consider the correlations between the forward and
reverse systematic errors.

\begin{figure}
\includegraphics[width=\columnwidth, viewport=0.5in 0in 7.25in 3.25in, clip]{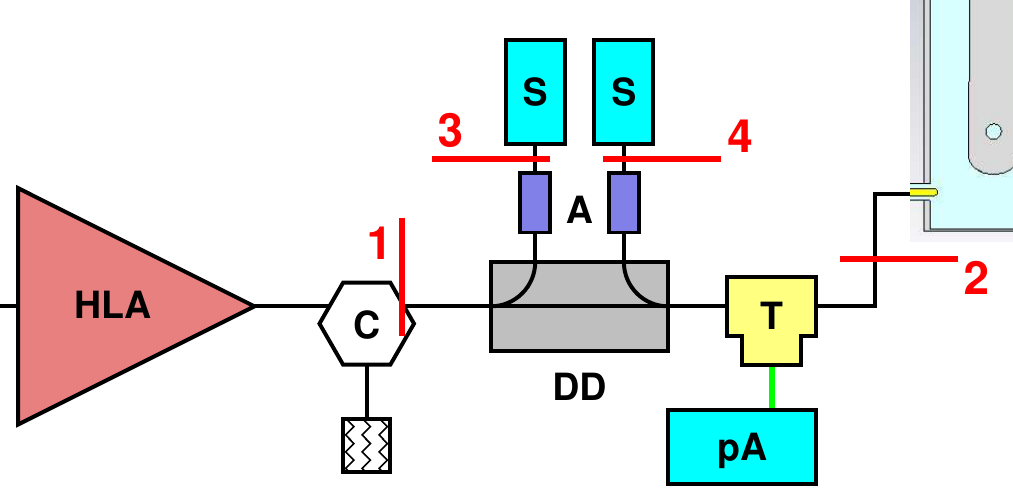}
  
\caption{Detail of the plasma processing RF system schematic.  Red
  lines: planes for measurements to characterize the directivity and
  mismatch between the dual directional coupler and the FPC\@.  A:
  attenuators; C: circulator with load; DD: dual-directional
  coupler; HLA: high-level amplifier; pA: picoammeter; S: power
  sensor; T: bias T.\label{F:syserr}}
\end{figure}

\Cref{F:syserr} shows a zoomed-in view of the RF system between the
high-power amplifier and the cavity, including the dual directional
coupler used to measure $P_f$ and $P_r$.  For the error analysis, we
did additional measurements on the portion of the circuit delimited by
the red lines.  The circulator is not included, as it is upstream of
the dual directional coupler, so that its mismatch does not adversely
affect $P_f$ and $P_r$ measurements (though the circulator mismatch
can produce ``ripple'' in the forward power reaching the cavity as the
drive frequency varies).  The bias T is included in the error
analysis, as it is downstream of the dual directional coupler and
hence its mismatch undermines the directivity of the measurements.

Referring to \cref{F:syserr}, we measure the forward scattering
parameters ($S_{21}, S_{31}, S_{41}$) with an incident forward wave at
Port 1 and the reverse scattering parameters
($S_{12}, S_{32}, S_{42}$) with an incident reverse wave at Port 2,
with a matched load to terminate unused ports.  We
consider the relationship between the forward and reverse amplitudes
at Plane 2 ($V_{f,2}, V_{r,2}$), representing the signals to and from
the cavity, and the signal amplitudes
($V_{fs} = V_3, V_{rs} = V_4$) reaching the power sensors.  In terms
of scattering parameters, we can write
\begin{eqnarray}
V_{fs} = V_3 &=& \frac{S_{31}}{S_{21}}\left(V_{f,2} + \varepsilon_f V_{r,2}\right)\\
V_{rs} = V_4 &=& S_{42}\left(\varepsilon_r V_{f,2} +  V_{r,2}\right)\, ,
\end{eqnarray}
where
\begin{eqnarray}
\varepsilon_f &\equiv& \frac{S_{32} S_{21}}{S_{31}}\\
\varepsilon_r &\equiv&  \frac{S_{41}}{S_{21} S_{42}}\, .
\end{eqnarray}
In the expressions for $\varepsilon_f$ and $\varepsilon_r$, the
coupling parameter is in the denominator and the ``leakage'' or
``cross-talk'' parameter is in the numerator.  The transmission
parameter ($S_{21}$) is present because we are expressing the coupling
port signals in terms of the forward and reverse amplitudes at Port
2.

When the cavity is connected, the cavity's reflection coefficient
$\Gamma_{cav} = V_{r,2}/V_{f,2}$ can be used to
express the forward and reverse signals in terms of the corresponding forward and
reverse amplitudes only, without cross terms:
\begin{eqnarray}
V_{fs} &=& \frac{S_{31}}{S_{21}}\left(1 + \Gamma_{cav} \varepsilon_f\right) V_{f,2}\\
V_{rs} &=& S_{42}\left(\frac{\varepsilon_r}{\Gamma_{cav}} + 1\right) V_{r,2}\, ,
\end{eqnarray}
Our scalar correction corresponds to an assumption of no 
cross-talk, so that $\varepsilon_f = \varepsilon_r = 0$.  Upper bounds on
the relative error due to this approximation are hence
$|\Gamma_{cav} \varepsilon_f|$ for $P_f$ and
$|\varepsilon_r/\Gamma_{cav}|$ for $P_r$.

\begin{figure}
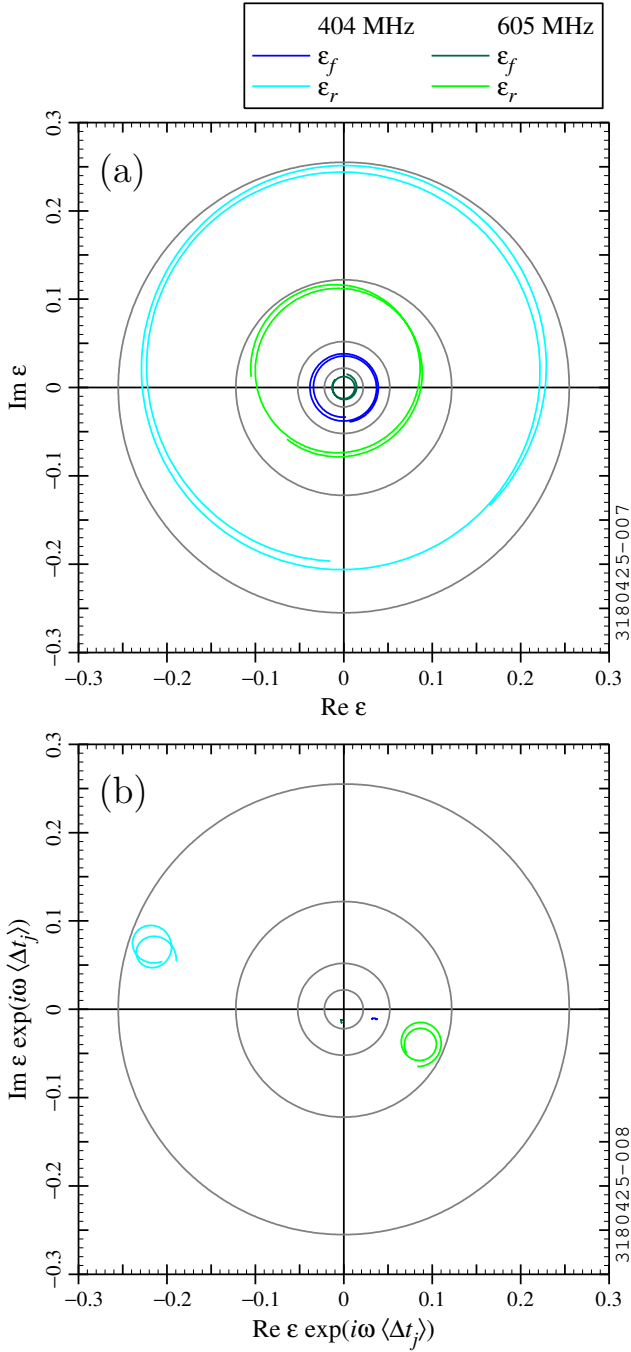

\GRAFoffset{-65}{-65}
\GRAFwidth[\sqwidth]{490}{560}
\GRAFlabelcoord{15}{355}
\incGRAFlabel{err_cmpr_mat_ri_circ_a}{(a)}\\[-3ex]
\GRAFwidth[\sqwidth]{490}{490}
\incGRAFlabel{err_cmpr_mat_ri_adj_circ_a}{(b)}\\[-3ex]
  
\caption{Relative error coefficients calculated from network analyzer
  measurements of the dual directional coupler, bias T, and associated
  cables.  (a) without delay adjustment; (b) with adjustment of the
  phase according to the mean delay ($\langle \Delta t_j\rangle$).
  Gray circles: magnitudes used in the systematic error
  analysis.\label{F:epsilons}}
\end{figure}

In general, the $S$-parameters are frequency-dependent, so the scalar
or vector correction should be frequency-dependent as well.  However,
for narrow-band measurements, frequency-independent scalar corrections
are generally used.  In our case, because we are shifting the drive
frequency by up to several MHz and often observe some ripple in the
signals as a function of frequency, we used a frequency-dependent
scalar correction in the final analysis of the measured values ($P_f$,
$P_r$, $P_t$).  However, this did not make much difference relative to
the other effects we have described.

In Ref. \cite{NIMA830:22to29}, a trombone phase shifter is used to
vary the phase difference between the dual directional coupler and the
cavity and gauge the impact of the phase on the measured signals.  For
our measurements, we swept the frequency to produce a similar effect.
\Cref{F:epsilons}a shows polar plots of $\varepsilon_f$ and
$\varepsilon_r$ obtained from measurements of the complex
$S$-parameters with a 20 MHz span; the phase change is about $4\pi$
radians over this frequency range.  In \cref{F:epsilons}b, the mean
delay for each of the measured $S$-parameters is subtracted.  With
this adjustment, the $\varepsilon_r$ values appear as approximate
circles, as would be expected for the sum of two signals with a varying
relative phase.  We interpret this to be the sum of a smaller direct
cross-talk signal from the forward wave to the reverse port of the
dual directional coupler and a larger signal reflected by downstream
mismatches (primarily from the bias T in our case) and coupled to the
reverse port.  The $\varepsilon_f$ values are significantly smaller in
magnitude than the $\varepsilon_r$ values, but still show some
frequency dependence, likely due to small reflections upstream of the
dual directional coupler.

The gray circles in \cref{F:epsilons} represent the ``worst-case''
values of $|\varepsilon_f|$ and $|\varepsilon_r|$ used for the error
analysis.  These values are listed in \cref{T:ErrVals}, along with
$|\Gamma_{cav}|$ values used for the various cases.  We note that
$|\Gamma_{cav}|$ includes the FPC cold window, rigid transmission
line, warm window, and RF adapters (though their contribution to the
return loss is small).
The systematic error estimates for $P_f$ and $P_r$ given in the text
are based on the values shown in \cref{T:ErrVals} after conversion
from amplitude to power.

\begin{table}[htbp]

\caption{Systematic error analysis parameters.\label{T:ErrVals}}
\centering
\begin{ruledtabular}
\begin{tabular}{lcc}
Frequency  & 404 MHz & 605 MHz \\ \hline
  $|\varepsilon_f|$ & $5.2 \cdot 10^{-2}$ & $2.2 \cdot 10^{-2}$ \\
  $|\varepsilon_r|$ & $2.55 \cdot 10^{-1}$ & $1.22 \cdot 10^{-1}$ \\
  $|\Gamma_{cav}|$ (plasma off) & 0.97 & 0.80 \\
  $|\Gamma_{cav}|$ (plasma on) & 0.97 & 0.98 \\ \hline
  $|\varepsilon_t|$ & $6.0 \cdot 10^{-2}$ & $1.0 \cdot 10^{-1}$ \\
\end{tabular}
\end{ruledtabular}

\end{table}

\subsection{Uncertainty cross-checks: \texorpdfstring{$P_f$}{Pf} and \texorpdfstring{$P_r$}{Pr}}

For the measurements described above, we terminated unused ports with
a matched load.  An alternative approach is to measure $S_{31}$ and
$S_{42}$ with different terminations (load, open, short) on the unused
Port 2 or Port 1 and subtract the results.  We did such measurements
and an associated analysis to obtain alternative values of
$\varepsilon_f$ and $\varepsilon_r$ as a cross-check.  The gray
circles in \cref{F:epsilons} correspond to the worst-case scenario
for both approaches (though we considered only the frequency range for
plasma measurements for the latter analysis).  This resulted in larger
circles for the $\varepsilon_f$ cases.

We note that the values of $\varepsilon_f$ and $\varepsilon_r$
correspond to an overall directivity that is significantly worse that
the directivity specification for our dual directional coupler: the
worse case of $\varepsilon_r = 2.55 \cdot 10^{-1}$ corresponds to a
directivity of about 12~dB, in contrast to the specified directivity
of 35~dB or higher.  This can be explained by the mismatch in the
system, as seen by reflection measurements ($S_{11}$ and $S_{22}$).  A
simple analysis indicates that $|\varepsilon_r|$ should be between
$|S_{11}|$ and $|S_{11}/(S_{21} S_{12})|$; our measured values are
consistent with this prediction.  Our reflection measurements are
roughly consistent with what we expect based on the bias T mismatch
specified by the manufacturer (SWR = 1.3).  Thus, the accuracy of the
measurements could be improved by mitigation of mismatches between the
dual directional coupler and the cryomodule.

\subsection{Uncertainty evaluation: \texorpdfstring{$P_t$}{Pt}}

Mismatch can adversely affect the measurement of $P_t$, as was the
case for $P_f$ and $P_r$.  Because the probe pickup antenna is weakly
coupled, we expect signals reflected from the RF measurement circuit
back toward the cavity to be reflected again toward the power sensor,
adding to or subtracting from the direct signal.  Hence we used
mismatch measurements to infer a worst-case relative error in $V_t$,
with error propagation for $P_t$ and $U$.  In this model, the
first-order worst-case relative error in $V_t$ is $\varepsilon_t$,
which is the product of the measured reflection $S$-parameters looking
``upstream'' toward the cavity pickup antenna and looking
``downstream'' to the RF measurement circuit for $P_t$.  The
corresponding $|\varepsilon_t|$ values, included in
\cref{T:ErrVals}, were used for the estimated systematic error in
$U^*$ discussed in \cref{S:th}.  As was the case for $P_f$ and
$P_r$, the bias T on the pickup line provides additional mismatch and
worsens the systematic errors.

\section{Power dissipation calculations}

The method for estimating the (total) power dissipation in the cavity
and the power dissipation in the plasma will be described in this
appendix, providing the basis for the corresponding results in
\cref{S:PfPdFreq}.

\subsection{Total power dissipation\label{A:Pd}}

The power dissipation $P_d$ in the cavity during steady plasma
processing can be obtained from the measured CW power values via a
direct calculation:
\begin{equation}
P_d = P_f - P_r - P_t
\end{equation}
The $P_t$ term, usually small compared to $P_f$ and $P_r$, is often
omitted.

\begin{figure}
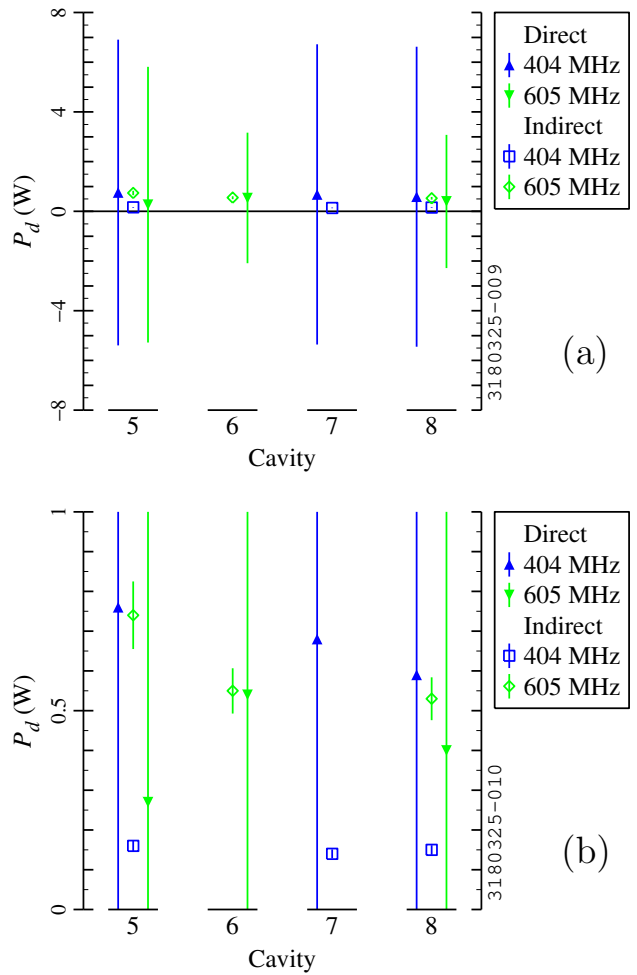

\GRAFoffset{-65}{-65}
\GRAFwidth[\landwidth]{490}{390}
\GRAFlabelcoord{360}{35}
\incGRAFlabel{pd_calc_bar_rev2_a}{(a)}\\[-2ex]
\incGRAFlabel{pd_calc_bar_rev2_a_zoom}{(b)}\\[-2ex]
  
\caption{(a) Calculated power dissipation during steady plasma processing
  with estimated errors: (a) zoomed-out view showing errors for the direct
  calculation; (b)
  zoomed-in view showing errors for the indirect calculation.\label{F:PdBars}}
\end{figure}

When the input coupler is poorly matched, this direct calculation of
$P_d$ is problematic, because $P_r \approx P_f$ and hence
$P_d \approx 0$.  This can be seen in \cref{F:PdBars}a: the solid
triangles indicate the calculated $P_d$ during steady plasma
processing for each cavity and mode combination.  Propagation of the
estimated errors in $P_f$ and $P_r$ (per \cref{A:errs}) leads to a
large uncertainty in $P_d$ and does not rule out negative $P_d$
values.

If $\Qext{1}$ and $\Qext{2}$ are known, we can calculate $P_d$ from
$P_f$ and $P_t$ without making use of $P_r$.  In case of a weakly
coupled pickup ($\Qext{2} \gg Q_0$), we can write
\begin{equation}
P_d \approx P_t \frac{\Qext{2}}{\Qext{1}} \left(2 \sqrt{\frac{\Qext{1}}{\Qext{2}}\cdot\frac{P_f}{P_t}} - 1\right)\,.\label{E:indir}
\end{equation} 
A derivation of the above equation can be found in a separate report
\cite{INDIR:REPORT}.  The hollow squares and diamonds in
\cref{F:PdBars}a show $P_d$ values calculated using \cref{E:indir}.
The error bars can been seen more clearly in the zoomed-in view shown
in \cref{F:PdBars}b.  The values are significantly different and the
uncertainties are significantly smaller when calculating $P_d$
indirectly.  The values listed in \cref{T:pfpd} are calculated using
the indirect method.  We note that our analysis assumes that
$\Qext{1}$ and $\Qext{2}$ are not perturbed by the plasma; we return
to this point in the next section.

\subsection{Power transfer to the plasma\label{A:Px}}

In \cref{S:PfPdFreq}, we provided estimates of the fraction of the
power dissipation in the plasma relative to the total power
dissipation.  These estimates are based on the assumption that the
input and pickup coupler strengths ($\Qext{1}, \Qext{2}$) and the
cavity's ``unperturbed intrinsic quality factor'' ($Q_{00}$) do not
change with plasma on.  We define $Q_{00}$ to be the quality factor
associated with the power dissipation in the cavity walls due to Ohmic
losses, not including the power dissipation in the plasma.  As the
plasma may perturb the field distribution of the resonant mode being
used to drive the plasma, the assumptions about constant $Q_{00}$ and
coupling strengths may not be exactly correct.  (In principle, a
measurement of the bandwidth with frequency sweep or decay time with
amplitude modulation after plasma ignition would provide a way to
infer $\Qext{1}$ and $\Qext{2}$ with the plasma on, but such a
measurement is not altogether straightforward.)  Though the answer may
not be exact, estimation of the power dissipated in the plasma is
nevertheless useful to get an idea of the process efficiency.

Assuming that $\Qext{1}$, $\Qext{2}$, and $Q_{00}$ are the same as
measured at low power without plasma, we use the measured value of
$P_t$ to infer $U$.  We infer the power dissipation in the cavity
walls from $U$ and $Q_{00}$.  Subtraction of this quantity from the
calculated (total) power dissipation $P_d$ gives the power $P_x$
dissipated in the plasma:
\begin{equation}
P_x = P_d -\frac{\omega U}{Q_{00}} = P_d - \frac{\Qext{2}}{Q_{00}} P_t\label{E:plasmaPd}
\end{equation}
where $\omega$ is the resonant (angular) frequency. Again,
$\Qext{2}$ and $Q_{00}$ are the values measured at low field without plasma.

\section{Plasma monitoring: RF power and pickup current\label{A:PowPick}}

\begin{figure*}
\GRAFlabelcoord{340}{1005}
\GRAFoffset{-65}{-65}
\GRAFwidth[\portwidth]{490}{1290}
\GRAFlabelbox{40}{30}%
\incGRAFboxlabel{cc_404_rf_bin_fdcor_pxc}{(a)} %
\incGRAFboxlabel{cc_605_rf_bin_fdcor_pxc}{(b)}\\[-2.5ex]

\caption{Measured and calculated RF values as a function of time with
steady plasma for (a) the 404 MHz mode and (b) the 605 MHz mode.\label{F:SteadyPows}}
\end{figure*}

\cref{F:SteadyPows} shows several quantities as a function of time with steady
plasma.  The forward power $P_f$ is measured. The stored
energy $U$ is inferred from the measured $P_t$ and the pickup coupling strength via
\begin{equation}
  \omega U = \Qext{1} P_t\, .
\end{equation}
The total power dissipation in the cavity ($P_d$) is calculated from
$P_f$, $P_t$, and the coupling strengths via the indirect method, per
\cref{E:indir}.  The estimated power dissipation in the plasma ($P_x$) is
calculated from the above quantities and the unperturbed 
intrinsic quality factor ($Q_{00}$) via \cref{E:plasmaPd}.  See
\cref{S:PfPdFreq} for further discussion of these results.  Some
upward spikes can be seen in $\omega U$ for Cavity 7 and Cavity 8;
these are unexplained.

The current collected by the pickup coupler during steady plasma is
shown in \cref{F:Ipick}, as discussed in \cref{S:lightI}.  As with
\cref{F:Ifpc}, the vertical lines in \cref{F:SteadyPows} and
\cref{F:Ipick} indicate time cuts of about 1 day or more.  Some
oscillation in the pickup current can be seen for the 404~MHz case
(Cavity~7, Round~5 and Cavity~5, Round~1); this is unexplained, but
could be due to non-ideal behavior of the picoammeter.

\begin{figure*}
\GRAFlabelcoord{15}{225}
\GRAFoffset{-65}{-65}
\GRAFwidth[\portwidth]{490}{390}
\GRAFlabelbox{40}{30}%
\incGRAFboxlabel{cc_404_rf_bin_fdcor_it_mixc_pi}{(a)} %
\GRAFlabelcoord{15}{255}
\incGRAFboxlabel{cc_605_rf_bin_fdcor_it_mixc_pi}{(b)}\\[-2.5ex]

\caption{Measured dc current collection by the pickup antenna as a function of time with
steady plasma for (a) the 404 MHz mode and (b) the 605 MHz mode.\label{F:Ipick}}
\end{figure*}

\bibliography{plas_v1}

\end{document}